\documentclass[12pt]{article}
\usepackage{amssymb,amsfonts,amsmath,latexsym}
\usepackage{color,float,fancyhdr}
\usepackage{graphicx}
\usepackage{wrapfig}
\usepackage{lscape}
\usepackage{rotating}
\usepackage{booktabs}
\usepackage[table]{xcolor}
\usepackage{url}
\usepackage{tabularx} %to have easy automated linebreaks within cells of the table

\usepackage[normalem]{ulem}%to strike out text (\sout{}

\usepackage[square,sort,comma,numbers]{natbib}
\setcitestyle{authoryear,open={(},close={)}}
\usepackage{changepage} % to move the table further to the left (to fit the page), i.e. to use "adjustwidth"

\usepackage[titletoc,title]{appendix} %for the appendix

\usepackage{setspace}
\usepackage{parskip}
\usepackage{etoolbox}

\usepackage[font={small,it}]{caption}

\newcommand{\subf}[2]{%
  {\small\begin{tabular}[t]{@{}c@{}}
   #1\\ #2
  \end{tabular}}%
}

\usepackage[latin1]{inputenc}
\usepackage{pstricks,pst-node,pst-text,pst-3d}

\usepackage[square,numbers]{natbib}

\usepackage{ifpdf}

\usepackage{todonotes}
\usepackage{color}

\newcommand{\R}{\mathbb{R}}
\newcommand{\Z}{\mathbb{Z}}

\newcommand{\p}{\mathbb{P}}


\def\R{\mathbb{R}}

\def\hQ{\widehat{Q}}
\newcommand{\1}{\mbox{\rm 1\hspace{-0.3em}I}}

\usepackage[top=1.5in, bottom=1.5in, left=1in, right=1in]{geometry}

\numberwithin{equation}{section}

\title{\textbf{\Large Pro-Cyclicality of Traditional Risk Measurements:\\Quantifying and Highlighting Factors at its Source\\[-1.1ex]}}
\author{Marcel Br\"autigam$\dagger\,\ddagger$, Michel Dacorogna*$\dagger$ and Marie Kratz$\dagger$\\[.1ex]
\small $\dagger$ ESSEC Business School, CREAR,  France \\[-1ex] \small * Prime Re Solutions, Switzerland\\[-1ex] \small $\ddagger$ Sorbonne University, LPSM \& LabEx MME-DII }

\date{November 2, 2019}

\begin{document}

\maketitle

\begin{abstract} 
%\vspace*{-2ex}
\noindent
Since the introduction of risk-based solvency regulation, pro-cyclicality has been a subject of concerns from all market participants. Here, we lay down a methodology to evaluate the amount of pro-cyclicality in the way financial institutions measure risk, and identify factors explaining this pro-cyclical behavior.\\
We introduce a new indicator based on the Sample Quantile Process (SQP, a dynamic generalization of Value-at-Risk), conditioned on realized volatility to quantify the pro-cyclicality, and evaluate its amount in the markets, considering 11 stock indices as realizations of the SQP. 
Then we determine two main factors explaining the pro-cyclicality: the clustering and return-to-the-mean of volatility, as it could have been anticipated but not quantified before, and, more surprisingly, the very way risk is measured, independently of this return-to-the-mean effect.\vspace*{1.5ex} \\
%
%\noindent{\emph {\emph {JEL} classification}: C13; C22; C52; C53; G01; G32} 
%
\end{abstract}

\newpage

\section{Introduction}
\label{sec-intro}
\vspace{-1ex}
%%%%%%%%%%%%%%%%%%%%%%%

% motivation
The introduction of risk based solvency regulations has brought the need for financial institutions (banks, insurance companies, ...) to evaluate their risk on the basis of probabilistic models. Since the RiskMetrics attempt by JP Morgan, the capital needed by companies to cover their risk is often identified to the quantile at a certain threshold of the return distribution of the portfolio (see~\citet{JPMorgan94}).
They called this risk measure Value-at-Risk (VaR), a name that has been adopted by the financial community. The question of the appropriateness of the risk measure to use for evaluating the risk of financial institutions has been heavily debated especially after the financial crisis of 2008/2009. For a review of the arguments on this subject, we refer e.g. to \citet{Chen18} and \citet{Ekt2015}. 

% literature and definitions to highlight diff between our def and economics def
Independently from the choice of adequate risk measures, there is an accepted idea that risk measurements are pro-cyclical: in times of crisis, they overestimate the future risk, while they underestimate it in quiet times. 
The term "pro-cyclicality" is used in the literature in various meanings. In economics, it is mainly related to the pro-cyclicality of economic agent behaviors in reaction to changing macro-economic conditions, thus reinforcing the macro-economic cycles. Here, we define this notion in a slightly different way. We adopt the point of view of risk management and regulation, by looking at the pro-cyclicality of risk measurements/estimations, which determine the capital amounts financial institutions are required to hold against their risk.
Of course, this would lead to a pro-cyclicality of economic agents behavior, namely they would not hold enough capital during quiet time to weather a crisis, whereas unnecessary capital  would be held during crises. 
Here, we do not look for modelling such behavior, but rather to evaluate the error made on risk measurement itself. It is thus a quite different point of view. This novel approach should complement the various economic studies performed on the topic.
Indeed, most of the literature we found on pro-cyclicality considers the economic approach.  Sometimes, the question of risk measurement is mentioned as a side remark within a macro-economic study (see e.g \cite{GordyHowells2006}), but without any attempt in quantifying the amount of under or over-estimation of the capital due to risk measurements. 

In the following, we briefly review the economic literature on the subject. 
%that is mostly concentrated on the former acception rather than the one we use in this study. We did not find papers directly tackling the problem we tackle here. It is mentioned sometimes as a side remark within a macro-economic study.

For a general review, we refer to \citet{Athanasoglou2014} and the references therein and also to \cite{Quagliariello08} for a review of empirical evidence of pro-cyclicality in banking behavior. In macro-economics, the general issue of pro-cyclicality has been analyzed with respect to the implications for banking regulation. In particular, there have been many discussions, already well before the 2008-09 crisis, on the fact that, in times of crisis, banks reduce lending to firms, accentuating then the liquidity crisis. It is thus not surprising that discussions have mostly focused on credit risk measurement. Economists have proposed models for explaining and countering this phenomenon of pro-cyclicality: Among them, \citet{White2006} reviews the possibilities of a new macro-financial stabilisation framework in which monetary and regulatory policies give more attention to avoiding the emergence of imbalances. 
\citet{Adrian10} focus on the behaviour of financial intermediaries (specifically: security broker dealers) showing empirically the pro-cyclicality of their leverage. Then,  in a subsequent study, they add as explanatory factor the behavior of the banks' Value-at-Risk (see \cite{Adrian13}). 
~\citet{Behn2016} use a quasi-experimental setup to show that model-based regulation in credit risk, which implies an increase in capital requirements, can increase the lack of credit. \citet{Gersbach2017} propose a simple two-sector economy model to examine the effects of counter-cyclical measures to fight against pro-cyclicality. 

Other authors examine the consequences of Basel II on this problem. For instance, \citet{KashyapStein2004} 
use simulation based arguments to conclude that the Basel II capital requirements have the potential to create additional pro-cyclicality in capital charges, while \citet{GordyHowells2006} argue that ``the marginal impact of introducing Basel II type regulations, depends strongly on the extent to which market discipline leads banks to vary lending standards pro-cyclically in the absence of binding regulation''. \citet{Repullo2008,Repullo2013} ``compare various bank capital regulation regimes using a dynamic equilibrium model of relationship lending'' and find that Basel II is more pro-cyclical than Basel I. Consequently, the issue of pro-cyclicality has been addressed explicitly in Basel III by creating a so called `counter-cyclical' capital buffer (see \citet{BaselIII}). Studies such as \citet{Repullo2011}, \citet{Rubio2016} and \citet{Llacay2019} deal with these new proposals in Basel III and review its consequences on the phenomenon.

% Description of our approach
Coming back to the perspective of risk management, there have been little attempts to study empirically on real financial data the pro-cyclicality of risk measure estimates themselves (among the few, see \citet{Bec2009}), and even less to quantify it.
Indeed, this pro-cyclicality is usually assumed to be a consequence of the volatility clustering and its return to the mean. Yet, we may question if this mean-reverting is fast enough to produce pro-cyclicality on a one year horizon, the typical time horizon used in risk management when measuring the risk. Hence the need to analyze further this effect.
Here, working with financial data, we are mostly interested in the way companies assess their need for capital. The amount of capital is computed as the VaR applied on probability distributions of future outcomes. The novelty of this paper is to propose alternative methods, on one hand, to quantify the predictive power of risk measurements, and, on the other hand, to explore the issue of pro-cyclicality by quantifying it, then  identifying some of its main factors. 

Such a study requires a dynamic reformulation of risk measurements departing from a pure static approach. While this approach holds for any risk measure in question, e.g. be it VaR or Expected Shortfall (ES), we focus on the example of VaR, which means departing from estimating quantiles of an {\it unconditional}  distribution. 
Clearly, different dynamic extensions already exist in the literature, see e.g. \cite{Engle2004}, \cite{Rossi2009}. 
Here, we focus on the simplest dynamic version of the `regulatory' risk measure VaR, %(while relying on a version of historical estimation).} 
%\sout{To do so, we generalize in a simple way the static `regulatory' risk measure VaR to a dynamic one.
namely the sample quantile process (SQP), considering the measurement itself as a random process. This notion has been proposed in the context of mathematical finance for options pricing, introduced by~\citet{Miura1992} in a general context while \citet{Akahori95} looked at it for Brownian Motion, and \citet{EmbSam1995} for the class of infinitely divisible stochastic processes. Using the SQP%with no specific underlying stochastic process
, we estimate it on financial time series and explore its dynamics and ability to predict correctly the future risk. 

We conduct our study on 11 stock indices (SI) of major economies to analyze the common behaviors and consider them as realizations of the generating process of financial returns, which we are exploring. 

We assume that the sample measurement is an estimation of the future risk of the time series, as commonly done in solvency regulations. We define a new indicator, a look-forward ratio that quantifies the difference between the historically predicted risk and the estimated realized future risk (measured {\it ex post}), to assess the accuracy of risk estimation.

Moreover, to take into account the market state, we choose the volatility as a proxy for qualifying it, since the realized volatility is high in times of crisis and low when there is little information hitting the market. 
That is why we analyze the look-forward SQP ratio conditioned to the realized volatility. As estimator for the realized volatility, and to ensure a good rate of convergence for it without requiring strong conditions on the underlying process, we use the mean absolute deviation instead of the root mean square deviation. We find a relatively strong (among the indices considered, between -46\% and -57\% for a VaR of 99\%) negative linear correlation between the log-ratio and the volatility, as well as a negative rank correlation of the same order (an average over all 11 indices of -49\% for the Spearman $\rho_S$). We find that in times of high volatility, the ratio is significantly below 1, while it can go even over 3, in times of low volatility, signaling a strong underestimation of future risk.

Looking for factors to characterize this effect, we proceed in two steps. First, we simulate identically and independently distributed (iid) random variables (rv's) and show that it produces negative correlation between the log ratio and the volatility (confirmed theoretically in \citet{Marcel2018}) as in real data but with less amplitude. Second, we model the stock indices log-returns by GARCH(1,1), a process that models the volatility clustering present in the data. Note that we use on purpose a simple GARCH to isolate the volatility clustering effect; we take first normal innovations, neglecting the tails, then Student-t innovations, to take them into account. The calibrated GARCHs are stationary and reproduce very well the average historical volatility. Using Monte Carlo simulations, we show that GARCH presents a very similar clustering behavior as in the data and gives also a strong negative correlation between the log-ratio and the volatility. This negative correlation is stronger than that obtained on the data. From the analysis of those two simple models, we conclude that the pro-cyclicality can be explained by two factors: (i) the way risk is measured as a function of quantiles estimated on past observations, and (ii) the clustering and return-to-the-mean of volatility. This opens a way to correct the risk measurement in order to better predict future risks.

The paper is organized in two main sections. In Section~\ref{sec-Qmethod}, we quantify the predictive power of risk measures. For this, in Subsection~\ref{ssec-dynRM} we formally treat the risk measure as a stochastic process by introducing the SQP. Then, we discuss its construction and explore its empirical behavior with 11 stock indices. In the next subsection, we define a new statistics, the look-forward SQP ratio, to explore the ability of the SQP to predict the risk for the future year.
In Section~\ref{sec-ProcCycl}, we analyze and quantify pro-cyclicality.  Conditioning the SQP on the realized annual volatility allows us to distinguish between the two main market states: normal versus crisis time. We show quantitatively that the SQP underestimates the risk in times of low volatility, while overestimating it in times of high volatility, confirming the pro-cyclicality of risk measurements. 
Then, looking for explanations, we identify two factors producing the (pro-cyclicality) effect with the help of two models, one with iid rv's, the other as a GARCH(1,1) process. We also discuss the influence of the sample size and the data frequency.
We summarize the results and discuss their consequences and future developments in the conclusion.

In this appendix we present three supplementary aspects. In Appendix~\ref{app:resid} we provide further evidence to conclude that the pro-cyclicality observed in the data is partially due to the GARCH effects (by analysing the residuals of the fitted GARCH process).
Then in Appendices~\ref{app:regr} and~\ref{app:vol_bins} we offer alternative ways of looking at the pro-cyclicality analysis of Section~\ref{ssec-volImpact}. Either via regression (see Appendix~\ref{app:regr}) or interpreting it more in line with the risk management point of view (see Appendix~\ref{app:vol_bins}).

%%%%%%%%%%%%%%%%%%%%%%%
\section{Quantifying the Predictive Power of Risk Measures}
\label{sec-Qmethod}
\vspace{-1ex}
%%%%%%%%%%%%%%%%%%%%%%%

\subsection{Risk Measure as a Stochastic Process}
\label{ssec-dynRM}
\vspace{-1ex}

Since the risk is changing with time, we choose to model it dynamically with a risk measure that is defined as a stochastic process. We use the simplest dynamic extension of the VaR, namely the Sample Quantile Process, first introduced in the context of options pricing (see~\citet{Miura1992},~\citet{Akahori95},~\citet{EmbSam1995}).  Here we do not assume any underlying model, and use directly this dynamic risk measure in its broad definition to check its performance in evaluating the risk on financial data.
\vspace{-2ex}

\subsubsection{Definitions}
\label{sssec-SQPDef}
\vspace{-2ex}

Let us quickly recall the definition of the VaR. For a loss random variable (rv) $L$ having a continuous distribution function $F_L$, the VaR at level $\alpha$ of $L$ is simply the quantile of order $\alpha$ of $L$:
\begin{equation}\label{eq:VaRDef}
VaR_{\alpha} (L) = \inf \Big\{ x : \p [L \le x] \ge \alpha \Big\} = F_L^{-1}(\alpha)
\end{equation}
where $F_L^{-1}$ denotes the generalized inverse function of $F_L$. The VaR is generally estimated on historical data, using the empirical quantile $F_{n}^{-1}(\alpha)$ associated with a $n$-loss sample $(L_{1}, \dotsc, L_{n})$  with $\alpha$ $\in~(0,1)$ defined by:
\begin{equation}\label{eq:empirical_VaRDef}
F_{n; L}^{-1}(\alpha) = \inf \Big\{ x : \frac{1}{n} \sum_{i=1}^{n} \1_{(L_{i} \leq x)} \geq \alpha \Big\}.
\end{equation}
Note that our choice of definition \eqref{eq:empirical_VaRDef} is such that the VaR equals an order statistics of the data. It would also be possible to obtain the VaR by linear interpolation between two order statistics. This might provide a slightly different value: the smaller the data set and/or the higher the threshold, the more the two quantile definitions usually differ. By construction and given our choice of definition of a quantile, the empirical quantile will then be larger or equal to that obtained by linear interpolation. Let us turn to the Sample Quantile Process. The loss is now represented by a stochastic process $L = (L_{t})_{t}$. The SQP $( Q_{\mu, \alpha, T,t}(L))_{ t \geq 0}$ of $L$ at threshold $\alpha$, with respect to a random measure $\mu$ defined on $\R^{+}$, is defined, for $0\leq T <t$, by (see e.g. \citet{Akahori95, EmbSam1995}):
\begin{equation}\label{eq:SQPDef}
Q_{\mu, \alpha, T, t}(L) = \inf \left\{ x : \frac{1}{\int_{t-T}^{t} d\mu(s)} \int_{t-T}^{t} \1_{(L_{s} \leq x)} d\mu(s) \geq \alpha \right\}.
\end{equation}
The SQP, denoted also by $(Q_{\mu, \alpha, T}(t))_{t\ge 0}=(Q_{\mu, \alpha, T,t} (L))_{t\ge 0}$ when no confusion is possible, is clearly a dynamic generalization of the VaR. Indeed, choosing $\mu$ in \eqref{eq:SQPDef} as the Lebesgue measure, corresponds to a rolling window VaR denoted by
\begin{equation}\label{eq:SQPDef2}
Q_{0,\alpha,T}(t) = \inf \left\{ x : \frac{1}{T} \int_{t-T}^{t} \1_{(L_{s} \leq x)}ds \geq \alpha \right\}.
\end{equation}
This can be read as the continuous version of the VaR defined in \eqref{eq:empirical_VaRDef} on the interval $[t-T,t)$.
Nevertheless, in practice, we estimate $Q_{0,\alpha,T}(t)$ for discrete observations $L_i$ of $L$, which comes back to considering a 'rolling window VaR'.

Our approach is to explore the SQP as a predictor of the risk, considering $\mu$ as a random measure. We focus on the particular case where  $\mu$ is defined by $\mu(s)=\mu_p(s)=|L_s|^p$, with $p \in \R$ : 
\begin{equation}\label{eq:SQPDef_specified_mu}
Q_{p,\alpha,T, t}=Q_{p,\alpha,T}(t)= \inf \left\{ x : \frac{1}{\int_{t-T}^{t} \lvert L_s \rvert^p ds} \int_{t-T}^{t} \1_{(L_{s} \leq x)} \lvert L_s\rvert^p  ds \geq \alpha \right\}.
\end{equation}
This choice of $\mu$ is arbitrary, but it is one of the simplest formulations that includes the loss, and gives back the Lebesgue measure when taking $p=0$, i.e. the VaR process with a rolling window. This latter case is quite important as it is what is used in practice by financial institutions; hence it will play the role of a benchmark in this study. This one of the traditional risk measure used with the expected shortfall to measure the risk of financial institution. Varying $p$ for our choice of $\mu$ is also a way to explore the impact of extreme movements: the higher $p$, the more statistical emphasis on the extremes. Looking at \eqref{eq:SQPDef_specified_mu}, we can see that positive values of $p$ will let the SQP put more weight on the tail of the distribution. And, correspondingly, negative values of $p$ will let the SQP put more weight on the center of the distribution. We concentrate on non-negative values of $p$ as we are interested in capturing and modelling the tail risk.

Throughout the paper, we consider the discrete approximation for the empirical estimator $(\hQ_{p,\alpha,T,t})_{t\ge 0}$ of the SQP, based on distinct (in our case daily) points in time $t_1,...,t_n$ and their sample $L_{t_1},...,L_{t_n}$, defined by
\begin{equation}\label{eq:SQPestim}
\hQ_{p,\alpha,T,t}= \inf \left\{ x : \frac{1}{\sum_{t_i \in [t-T,t) } \lvert L_{t_i} \rvert^p} \sum_{ t_i \in [t-T,t)} \1_{(L_{ t_i} \leq x)} \lvert L_{t_i} \rvert^p \geq \alpha \right\},
\end{equation}
and evaluate \eqref{eq:SQPestim} at subsequent points in time $t$ corresponding to a monthly rolling window.

\subsubsection{Empirical Exploration of the SQP}
\label{sssec-empExplor}
\vspace{-2ex}

Given this formalism, we are now going to build realized time series of SQP's computed over various stock indices, which we consider as various sample paths of the process. Since our goal is to look at the appropriateness of the VaR calculations done by financial institutions, it seems natural to use stock indices and see how our various SQP's perform on this data.

\paragraph{Data -}
\label{parag-data}

We consider data from 11 different stock indices which we have retrieved mainly through Bloomberg (and, for the time periods which were not available, through other global finance sources). The data used are the daily closing prices from Friday, January 2, 1987 to Friday, September 28, 2018. Detailed information about the countries and indices used can be found in Table~\ref{table:Indices_Info}. The reason for such a large set of indices is the need to see if the process has similar behavior over most of the developed economies. Moreover, we want a relatively long dataset, which is offered by all these stock indices.

As commonly done, we refer to the S\&P 500 index as our main example and focus on this specific index whenever showing figures. All the results for the other indices are available upon request and, if not commented on otherwise, show the same characteristics.
For the quantitative evaluation we will usually show both, the (average) results for each index (considering it as one realization of a stochastic process) as well as the average over all 11 indices with the corresponding standard deviation over the 11 results of each index.

For any time series, let us denote the closing price at time $t$ by $S(t)$ and by $\Delta t$ the interval between two consecutive prices.  We focus on daily intervals. We then define the log-return $X_{\Delta t}(t)$ by:
\begin{equation} \label{eq:ret}
X_{\Delta t}(t) = \ln \left (\frac{S(t)}{S(t-\Delta t)}\right) ~~.
\end{equation}
If not specified otherwise, by abuse of notation, we will refer to $X_i = X_{\Delta t} (t_i)$ as the daily log-return (with $\Delta t =1$ day) at time $t_i$. As the risk measures \eqref{eq:VaRDef} and \eqref{eq:SQPDef} are based on the losses, it is, in some situations, of interest to work with the log-return losses instead of the log-returns, simply defined as: $L_i = -X_i$, the daily log-return loss at time $t_i$.

\begin{table}[H]
\caption{\sf\small For each index we provide the country and the country code which the index represents, the Bloomberg index which is the name under which one can find the index on Bloomberg as well as a short description of the index.}
\footnotesize
\begin{tabular*}{\textwidth}{|p{2.4cm} p{2.2cm} p{2.8cm} p{7.4cm} |}
%\begin{tabularx}{\textwidth}{|l l l X|}    
\hline
 Country & Country Code & Bloomberg Index & Description %\\[0.5ex]
\\  \hline\hline 
Australia     & AUS &  AS30  & The All Ordinaries Index (AOI) is made up of the 500 largest companies listed on the Australian Securities Exchange (ASX) \\ % 
Canada        & CAN & SPTSX & The Canada S\&P/TSX Composite Index comprises around 70\% of market capitalization for Toronto Stock Exchange listed companies\\
France        & FRA &  SBF250  & The CAC All-Tradable Index is a French stock market index representing all sectors of the French economy (traded on Euronext Paris) \\
Germany       & DEU &  CDAX  & CDAX is a composite index of all stocks traded on the Frankfurt Stock Exchange in the Prime Standard and General Standard market segment \\
Great Britain & GBR &  ASX  & The FTSE All-Share Index (ASX) comprises more than 600 (out of 2000+) companies traded on the London Stock Exchange \\
Italy         & ITA & ITSMBCIG  & This index is called Comit Globale R it includes all the shares listed on the MTA and is calculated using reference prices \\
Japan         & JPN &  TPX & The Tokyo Price Index (TPX) is a capitalization-weighted price index of all First Section stocks (1600+ companies) of the Tokyo Stock Exchange \\
Netherlands   & NLD &   AEX   & The AEX index is a stock market index composed of a maximum of 25 of the most actively traded shares listed on Euronext Amsterdam \\
Singapore     & SGP &   STI   & FTSE Straits Times Index (STI) tracks the performance of the top 30 companies listed on the Singapore Exchange \\
Sweden        & SWE & OMXAFGX & Sweden OMX Aff\"arsv\"arldens General Index is a stock market index designed to measure the performance of the Stockholm Stock Exchange \\
United States of America & USA & SPX & S\&P 500 is an American stock market index based on 500 large companies listed on the NYSE or NASDAQ (around 80\% market capitalization)  \\
\hline
%\end{tabularx}
\end{tabular*}
\label{table:Indices_Info}
\end{table}

\paragraph{Empirical Study and Discussion -} 
\label{parag-study}

\begin{figure}[H]
\hspace*{1.5cm}
\begin{minipage}{0.5\textwidth}
\includegraphics[width=6cm,height=6cm]{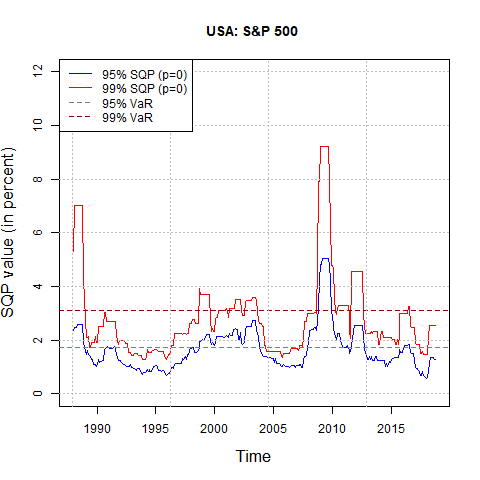}
\end{minipage}
\hspace{-1cm}
\begin{minipage}{0.5\textwidth}
\includegraphics[width=6cm,height=6cm]{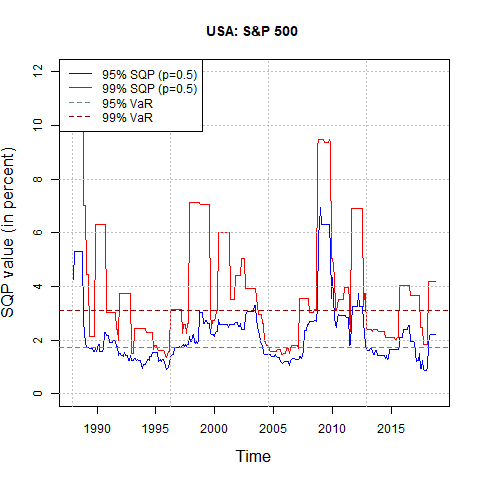}
\end{minipage}
\hspace*{1.5cm}
\begin{minipage}{0.5\textwidth}
\includegraphics[width=6cm,height=6cm]{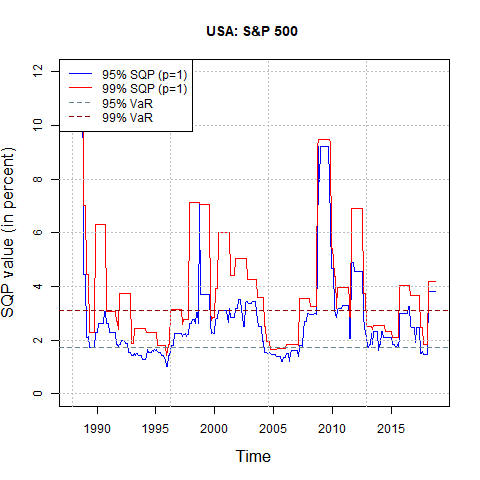}
\end{minipage}
\hspace{-1cm}
\begin{minipage}{0.5\textwidth}
\includegraphics[width=6cm,height=6cm]{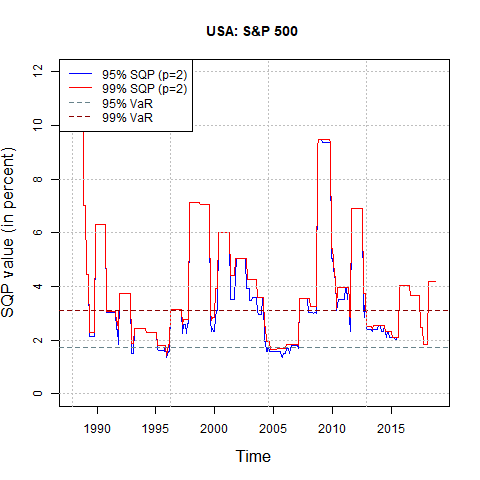}
\end{minipage}
\caption{\label{fig:SQRolling}\sf\small Sample Quantile Processes (SQP) rolling every month with $T = 1$ year, and  thresholds $\alpha=95$\% and 99\%, respectively. From left to right and from top to bottom: $p=0$, $p=0.5$, $p=1$ and $p=2$. Dashed lines correspond to VaR($\alpha$) computed over the whole sample. The y-scale is the same in the four graphs to allow for comparison between the different measures.}
\end{figure}

At first, we focus on a rather qualitative evaluation, illustrated on the S\&P 500. 
In Figure~\ref{fig:SQRolling}, we plot the sample paths of SQP at level $\alpha$ at monthly frequency and the unconditional VaR (i.e. computed on the overall sample) at the same level $\alpha$, as it is the regulatory benchmark. Note that we use in this study  the SQP for $p=0$, which is the rolling-window VaR, as our dynamic benchmark of risk. We look at the SQP for different parameter values $p$ of the random measure $\mu=\lvert L_S \rvert^p$: $p=0,0.5,1,2$. The sample size, $T$, is taken to be one year and we consider thresholds $\alpha= 95$\% and 99\%, respectively.

When $p = 0.5$, the values of the SQP are not only more volatile ({\it i.e.} with a wider range) than for $p=0$, but also bigger than the SQP for $p=0$, in particular with extreme values that seem to have increased notably. Hence, the SQP, with $p=0.5$, is a measure that enables a larger discrimination between the losses.

For $p = 1$, the properties found for $p = 0.5$ are also present and accentuated.  The volatility of the values of the SQP has once again increased for $\alpha = 0.95$. The range of the SQP is also bigger than for $p<1$. As for $p = 0.5$, the SQP for $p=1$ is a measure that enables a bigger discrimination between the losses. But we can also see that the differences between the SQP for the two thresholds has clearly diminished for extreme values. To conclude, the SQP for $p = 1$ is close to the SQP for $p = 0.5$, apart for the most extreme values at threshold $\alpha =0.95$, and continues the trend seen for the latter. In addition, the SQP for $p = 2$ is very close to the previous SQP for $p=1$. However, the differences between the SQP with the different thresholds are extremely small and almost vanished. Hence, it is difficult, if not impossible, to distinguish between the 95\% and 99\% thresholds. The 95\% threshold collapses in fact on to the 99\% threshold. 

As can be seen on the last graph, both SQPs hover around the VaR(99\%). Moreover, for the 99\% threshold, the SQP with $p=2$ equals the SQP with $p=1$, which does not make it interesting anymore. We then conclude that the choice of $p=1,2$ to define the SQP may not be the best one to assess the intensity of losses.

In summary, we can observe the following. For a fixed threshold $\alpha$, the SQP increases with higher powers of $p$. Also, the higher the parameter $p$, the more volatile the SQP gets. At the same time, the difference of the SQPs at level $\alpha = 95 \%$ and 99\% diminishes with increasing $p$ - we could observe that, for $\alpha = 99\%$, the SQP for p=1 and p=2 coincide. From the figures, $p=0.5$ seems optimal for tail discrimination.

These observations can also be confirmed when looking at it more quantitatively as done in Table~\ref{tbl-avSQP} and its figure on the right for the S\&P 500, as well as for all indices in Table~\ref{tbl-avSQPAll}. In Table~\ref{tbl-avSQP}, we compare the average values achieved by the various monthly rolling SQP's with sample size $T$ equal to one year, as a function of the parameter $p$ and the threshold $\alpha$ for the S\&P 500. We notice from the figure again that for the highest thresholds (above and including 99\%), the values for $p=1$ and $p=2$ are the same. We see in Table~\ref{tbl-avSQPAll}, where we consider the behavior of the 11 different indices as well as their average, that there is no marked difference between the various indices. The behavior is the same, the higher 

\begin{table}[H]
%\hspace*{1cm}
\begin{minipage}{0.62\textwidth}
{\footnotesize
\begin{tabular}{lccc}\hline
\\[-1.5ex]
Risk measures & $\alpha$=95\% &  $\alpha$=99\% & Ratio\\
&&&  SQP$(99\%)$/SQP$(95\%)$ \\[0.5ex]
\hline
\hline
\\[-1.5ex]
~~SQP (p=0)         &   1.65\%   &   2.80\%   &     1.70  \\[0.5ex]
~~{\bf SQP (p=0.5)} &{\bf 2.19\%}&{\bf 4.29\%}&{\bf 1.96} \\
~~SQP (p=1.0)       &   3.24\%   &   4.43\%   &     1.37  \\
~~SQP (p=2.0)       &   4.30\%   &   4.43\%   &     1.03  \\[0.5ex]
\hline
\end{tabular}
}
\end{minipage}
\begin{minipage}{0.3\textwidth}
\includegraphics[width=6cm,height=6cm]{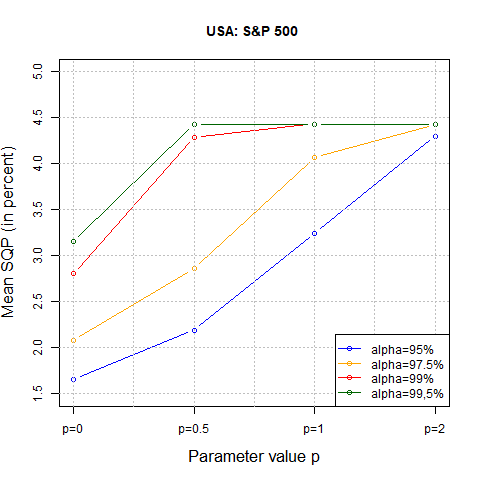}
\end{minipage}
\caption{\label{tbl-avSQP}\sf\small Average over the whole sample of rolling-window SQPs with sample size $T=1y$ (S\&P 500) as a function of the power $p$ for various thresholds $\alpha$.}
\end{table}
\begin{table}[H]
\begin{center}
\scriptsize
\parbox{440pt}{\caption{\label{tbl-avSQPAll}\sf\small Average over the whole sample of the rolling-window SQPs with sample size $T=1y$ for each of the 11 stock indices (in \%) and associated ratio of the SQPs between two thresholds (95 and 99\%) - using historical data. In the last column, we present the average over all indices $\pm$ the standard deviation over the 11 displayed values.}\vspace{-2ex}}
\begin{tabular*}{440pt}{l c c c c c c c c c c c c}    \hline \\[-1ex]
Mean for & AUS & CAN& FRA& DEU & ITA & JPN & NLD & SGP &SWE & GBR & USA & {\bf AVG ($\pm~\sigma$)} \\[1ex] \hline\hline
	\\[-1ex]
	& \multicolumn{11}{c}{Mean 1y  in \% ($\alpha= 95 \%$)}  \\	\cline{2-13}\\ [-1.5ex]
$p=0$   & 1.44 & 1.44 & 1.91 & 1.92 & 2.08 & 1.97 & 2.00 & 1.77 & 1.99 & 1.58 & 1.65  & {\bf 1.80 (0.23)} \\ 
$p=0.5$ & 1.92 & 1.96 & 2.56 & 2.56 & 2.69 & 2.68 & 2.71 & 2.61 & 2.64 & 2.11 & 2.19 & {\bf 2.42 (0.31)} \\
$p=1$   & 3.16 & 2.68 & 3.15 & 3.51 & 3.51 & 3.82 & 3.39 & 4.05 & 3.33 & 2.73 & 3.24 & {\bf 3.33 (0.41)} \\
$p=2$   & 4.11 & 3.65 & 4.35 & 4.32 & 4.76 & 5.25 & 4.49 & 5.26 & 4.61 & 3.58 & 4.30 & {\bf 4.43 (0.55)} \\[2ex]
 & \multicolumn{11}{c}{Mean 1y in \% ($\alpha= 99 \% $)}  \\	\cline{2-13}\\[-1.5ex]
$p=0$   & 2.37 & 2.53 & 3.18 & 3.16 & 3.41 & 3.36 & 3.37 & 3.35 & 3.24 & 2.69 & 2.80 & {\bf 3.04 (0.38)} \\ 
$p=0.5$ & 4.07 & 3.58 & 4.27 & 4.21 & 4.72 & 5.24 & 4.38 & 5.18 & 4.53 & 3.49 & 4.29& {\bf 4.36 (0.56)} \\
$p=1$   & 4.24 & 3.73 & 4.49 & 4.51 & 4.96 & 5.39 & 4.67 & 5.40 & 4.82 & 3.67 & 4.43 & {\bf 4.57 (0.57)} \\
$p=2$   & 4.24 & 3.73 & 4.49 & 4.51 & 4.96 & 5.39 & 4.67 & 5.40 & 4.82 & 3.67 & 4.43 & {\bf 4.57 (0.57)} \\[2ex]
 & \multicolumn{11}{c}{Ratio of Mean 1y at 99\% over 95\%}  \\	\cline{2-13}\\[-1.5ex]
$p=0$   & {\it ~1.65} & {\it ~1.76} & {\it ~1.66} & {\it ~1.65} & {\it ~1.64} & {\it ~1.71} & {\it ~1.68} & {\it ~1.89} & {\it ~1.63} & {\it ~1.70} & {\it ~1.70} & \textbf{\textit{~1.70 (0.07)}} \\ 
$p=0.5$ & {\it ~2.12} & {\it ~1.83} & {\it ~1.67} & {\it ~1.64} & {\it ~1.75} & {\it ~1.96} & {\it ~1.62} & {\it ~1.98} & {\it ~1.72} & {\it ~1.65} & {\it ~1.96} & \textbf{\textit{~1.81 (0.17)}} \\
$p=1$   & {\it ~1.34} & {\it ~1.39} & {\it ~1.43} & {\it ~1.28} & {\it ~1.41} & {\it ~1.41} & {\it ~1.38} & {\it ~1.33} & {\it ~1.45} & {\it ~1.34} & {\it ~1.37} & \textbf{\textit{~1.38 (0.05)}} \\
$p=2$   & {\it ~1.03} & {\it ~1.02} & {\it ~1.03} & {\it ~1.04} & {\it ~1.04} & {\it ~1.03} & {\it ~1.04} & {\it ~1.03} & {\it ~1.05} & {\it ~1.03} & {\it ~1.03} & \textbf{\textit{~1.03 (0.01)}} \\[2ex]
\hline
\end{tabular*}
\end{center}
\vspace{-2ex}
\end{table}

the value of $p$, the larger the average (in absolute terms). Nevertheless, such assessment has to be put in perspective when taking the standard deviation around the average value into account: we can observe that there is a difference in the behavior of the ratio between $p=0$ and $p>0$, especially visible for $p=99\%$. In the same table, the ratio between the two SQP averages at different thresholds $\alpha$ is the largest on average for $p=0.5$ confirming our remarks above about tail discrimination (this is not true for all indices; particularly the European ones: DEU, NLD, GBR do not follow this behavior).

\subsection{Predictive Power for the Risk using the SQP}
\label{ssec-prediction}
\vspace{-1ex}

After having introduced and analyzed the SQP for various measures $\mu$, we now focus on the main goal namely testing the SQP's ability to assess the estimated future risk correctly - what we call predictive power. Let us briefly point out why this differs from the classical backtesting, which designates a statistical procedure that compares realizations with forecasts. First, recall that, for the VaR, some of the backtesting methods used by European banks and the Basel committee are based on the so-called violation process counting how many times the estimated/forecast VaR($\alpha$) has been violated by realized losses during the following year; those VaR 'violations' should form a sequence of iid Bernoulli variables with probability $(1-\alpha)$ (see e.g. \citet{Christoffersen1998} for simple binomial test, \citet{kratz2018} for multinomial one).
Here instead, we want to assess the risk, thus the quality of estimation. While the violation ratio in \cite{Danielsson2002} gives a measure of the quality of the estimation (based on the violation process), here we do not question at all if the violation process is in line with the acceptable number of exceptions. The analysis is based on the returns time series itself and the goal is rather, when considering the risk estimation as a stochastic process, to quantify precisely how well or wrong our VaR estimate performed, compared to the realized future value of the VaR. It is an alternative and new way of looking at risk, which, subsequently, leads also to another way to assess the quality of the risk estimation. Moreover, this way of risk quantification can be applied to any other risk measure besides the VaR (as for instance to the Expected Shortfall).

\subsubsection{Look-forward SQP Ratio}
\label{sssec-SQPRatio}
\vspace{-2ex}

To define a look-forward ratio, we consider SQP estimators $(\hQ_{p,\alpha,T}(t))_t$, computed according to \eqref{eq:SQPestim} during a period (of length T years) that ends at time $t$;  for ease of notation, we omit the hat above in the following.

We measure the quality of the risk estimation by comparing it to the VaR, estimated one year later $(p=0, T=1)$, i.e. $Q_{0, \alpha, 1}(t+1y)$. It means to check if the estimate $Q_{p,\alpha,T}(t)$ is a good estimate at time $t$ of the unknown future value $Q_{0, \alpha,1}(t+1y)$. Hereby, it is important to mention that the reference value, the VaR one year later $Q_{0, \alpha, 1}(t+1y)$, will always be computed over a period of one year ($T=1$) as it should represent the realized risk during that period of time. Note also that we compare all versions of the SQPs to the VaR, as this is the reference risk measure asked by many regulators.

This leads us to introduce a new rv, the ratio of SQPs, denoted by $R_{p, \alpha,T}(t)$, which quantifies the difference between the historically predicted  risk $Q_{p,\alpha,T}(t)$ and the estimated realized future risk $Q_{0, \alpha,1}(t+1y)$ computed using the period of time $[t,t+1y)$:

\begin{equation} \label{eq:VaRRatio}
R_{p, \alpha,T}(t) = \frac{Q_{0, \alpha, 1}(t+1y)}{Q_{p,\alpha, T}(t)}.
\end{equation}

To be clear, let us look at a simple example. Choosing $\alpha = 95\%$, $t=$ January 2014, and $T=1$ year, the denominator $Q_{p,95\%,1}(t)$ of the ratio $R_{p,95\%,1}(t)$ corresponds then to the empirical estimate of the SQP computed on a 1 year sample from January 2013 ($=t-T$) to December 2013 ({\it i.e.} the month before $t$, as we are considering the interval $[t-T,t)$), which acts as an estimate for the future risk in 2014  (if $T$ would be 3 years, $Q_{p,0.95, 3}$ would be computed from January 2011 up to December 2013). The numerator corresponds to the future risk $Q_{0, \alpha, 1}(t+1y)$ and is defined as the rolling VaR estimated on the one year sample from January 2014 ($t$) until December 2014 ({last month before $t+1y$).

Consequently, if we look at the values of the ratio $R_{p,\alpha,T}(t)$, a value near 1 means that the prediction $Q_{p,\alpha,T}(t)$ correctly assesses the future risk estimated by $Q_{0, \alpha,1}(t+1y)$. Similarly, if the ratio is above 1, we can say that we have an under-estimation of the future risk, i.e. the observed future risk. And to the contrary, if it is below 1, we have an over-estimation of the future risk, which would lead to an over-evaluation of the needed capital.

\subsubsection{Assessing the Average SQP ratio}
\label{sssec-avSQPR}
\vspace{-2ex}

Taking a monthly rolling window for the ratios $R_{p,\alpha,T}(t)$ (i.e. updating $t$ every month), we compute the average SQP ratio over the whole sample. Here we focus on looking at the overall averages (over the 11 different indices) of the monthly SQP ratios. But results for the 11 different indices on their own are reported in Table~\ref{tbl-avSQPR}, too.

\begin{table}
\begin{center}
\parbox{430pt}{\caption{\label{tbl-avSQPR}\sf\small SQP ratios (defined in~\eqref{eq:VaRRatio}) average over the whole historical sample, for 1 year, for each index and for two different thresholds (95 and 99 \%). In the last column, we present the average over all indices $\pm$ the standard deviation over the 11 displayed values.}\vspace{-2ex}}
\scriptsize
\begin{tabular*}{430pt}{l c c c c c c c c c c c c}    \hline \\[-1ex]
Mean for & AUS & CAN& FRA& DEU & ITA & JPN & NLD & SGP &SWE & GBR & USA & {\bf AVG ($\pm~\sigma$)} \\[1ex] \hline\hline
	\\[-1ex]
	& \multicolumn{11}{c}{SQP ratio for 1y ($\alpha= 95 \%$)}  \\	\cline{2-13}\\ [-1.5ex]
$p=0$    &  1.02 & 1.06 & 1.06 & 1.09 & 1.07 & 1.08 & 1.07 & 1.05 & 1.07 & 1.06 & 1.05 & \bf{1.06 (0.02)} \\ 
$p=0.5$ & 0.79 & 0.81 & 0.79 & 0.83 & 0.82 & 0.82 & 0.81 & 0.78 & 0.83 & 0.82 & 0.81 & \bf{0.81 (0.02)} \\
$p=1$   &   0.63 & 0.64 & 0.64 & 0.63 & 0.64 & 0.66 & 0.65 & 0.60 & 0.66 & 0.66 & 0.65 & \bf{0.64 (0.02)} \\
$p=2$   & 0.49 & 0.47 & 0.49 & 0.50 & 0.48 & 0.48 & 0.49 & 0.44 & 0.49 & 0.50 & 0.49 & \bf{0.48 (0.02)} \\[2ex]
 & \multicolumn{11}{c}{SQP ratio for 1y ($\alpha= 99 \%$)}  \\	\cline{2-13}\\[-1.5ex]
$p=0$   &1.02 & 1.08 & 1.04 & 1.08 & 1.08 & 1.11 & 1.09 & 1.11 & 1.07 & 1.06 & 1.07 & \bf{1.07 (0.03)} \\ 
$p=0.5$ & 0.79 & 0.81 & 0.83 & 0.86 & 0.81 & 0.82 & 0.86 & 0.81 & 0.81 & 0.86 & 0.82 & \bf{0.83 (0.02)} \\
$p=1$   & 0.73 & 0.76 & 0.77 & 0.78 & 0.76 & 0.77 & 0.78 & 0.74 & 0.74 & 0.79 & 0.77& \bf{0.76 (0.02)} \\
$p=2$   & 0.73 & 0.76 & 0.77 & 0.78 & 0.76 & 0.77 & 0.78 & 0.74 & 0.74 & 0.79 & 0.77& \bf{0.76 (0.02)} \\[2ex]
\hline
\end{tabular*}
\end{center}
\vspace{-2ex}
\end{table}

We see, in Table~\ref{tbl-avSQPR}, that all the ratios are clearly different from 1 (taking also into account the standard deviation around the average value), telling us that the future risk is, on average, not well predicted. Furthermore, we clearly see for both choices of $\alpha$ that the average ratio is above 1 for $p=0$ - even when considering two standard deviations around the average value. We find back the well known result that the historical estimation of VaR underestimates the risk (on average).
For $p>0$, we see that the values are below 1 and clearly decrease with $p$ for both $\alpha=95\%$ and $99\%$ (apart from $p=2$ for $\alpha = 99\%$). This implies that our choice of parameter $p$ puts too much weight to the tails on average, causing a clear overestimation of the actual risk.

When comparing the average ratio for our two thresholds $\alpha$, we see that it does not change a lot for $p=0,0.5$, while the ratio increases with $\alpha$ for $p=1,2$. This increase is not surprising as for those values of $p$, more weight is added to the tail, so the larger the threshold $\alpha$, the smaller the difference to the actual future risk, and thus the smaller the overestimation. 

To complete this analysis, we look at the Root Mean Square (RMS) distance and compute the associated Root Mean Square Error (RMSE), i.e. the average RMS distance of the ratio from a perfect prediction (i.e. ratio of 1):
\begin{equation}\label{eq:rmse}
RMSE := \sqrt{\frac{1}{N} \sum_{j=1}^N \left(R_{p, \alpha, T}( t_j) - 1\right)^2}, 
\end{equation}
where $t_1,...,t_N$ are the $N$ time points the ratio is evaluated at. Recall that, as we are working with a monthly rolling window, $N$ equals 12 times the amount of years on which the ratio is tested on. This RMSE measures the "error" of the forecasted risk for a certain sample path.
\begin{figure}[H]
\centering
\begin{minipage}{0.6\textwidth}
\includegraphics[width=7cm,height=7cm]{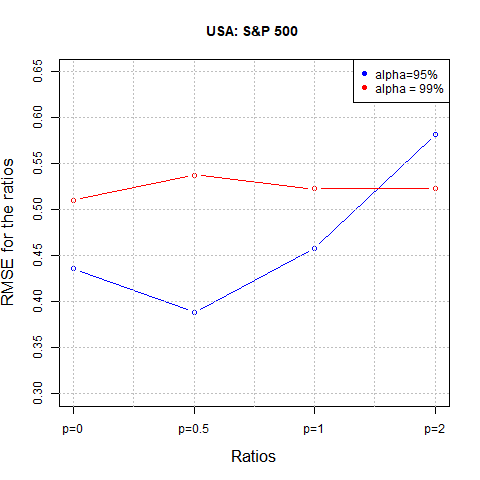}
\end{minipage}
\hspace{-2.5cm}
\begin{minipage}{0.4\textwidth}
\parbox{7cm}{\caption{\label{fig:RMSE-ratios} \sf\small RMSE (defined in \eqref{eq:rmse}) of the SQP Ratios $R_{p, \alpha,1}(t)$ taking $p=0,0.5,1,2$ and two thresholds $\alpha = 95\%$ (in blue), $99\%$ (in red), computed on the S\&P 500}}
\end{minipage}
\end{figure}

\vspace{-3ex}
For the S\&P 500 index, we see in Figure~\ref{fig:RMSE-ratios} that the error increases with the threshold: it is smaller for 95\% than for 99\%, except for the SQP measure with p=2. For the 99\% threshold, the error increases slightly from $p=0$ to $p=0.5$ where it has its maximum, and levels slightly off for $p=1,2$. We observe a different, almost opposite pattern in the case of 95\%: it decreases from $p=0$ to its minimum for $p=0.5$, then increases strongly with $p$. All the values are clearly above 0 and close to 0.5, suggesting that the various measures do not forecast well the future risk. 

\begin{table}
\begin{center}
\parbox{430pt}{\caption{\label{tbl-SQPratios-RMSE}\sf\small Average RMSE for the SQP ratios (eq.~\eqref{eq:VaRRatio}, with $T=1$ year) - using historical data for each index. In the last column, we present the average over all indices $\pm$ the standard deviation over the 11 displayed values.}\vspace{-2ex}}
\scriptsize
\begin{tabular*}{430pt}{l c c c c c c c c c c c c}    \hline \\[-1ex]
Mean for & AUS & CAN& FRA& DEU & ITA & JPN & NLD & SGP &SWE & GBR & USA & {\bf AVG ($\pm~\sigma$)} \\[1ex] \hline\hline
	\\[-1ex]
	& \multicolumn{11}{c}{RMSE of the SQP ratio for 1y ($\alpha= 95 \%$)}  \\	\cline{2-13}\\ [-1.5ex]
$p=0$   &  0.36 & 0.45 & 0.44 & 0.50 & 0.48 & 0.54 & 0.51 & 0.46 & 0.47 & 0.43 & 0.44& \bf{0.46 (0.05)} \\ 
$p=0.5$ & 0.34 & 0.42 & 0.39 & 0.43 & 0.40 & 0.46 & 0.43 & 0.42 & 0.40 & 0.38 & 0.39& \bf{0.40 (0.03)} \\
$p=1$   & 0.45 & 0.48 & 0.45 & 0.48 & 0.46 & 0.52 & 0.47 & 0.51 & 0.45 & 0.44 & 0.46& \bf{0.47 (0.03)} \\
$p=2$   & 0.57 & 0.59 & 0.56 & 0.56 & 0.56 & 0.61 & 0.56 & 0.62 & 0.57 & 0.55 & 0.58 & \bf{0.57 (0.02)} \\[2ex]
 & \multicolumn{11}{c}{RMSE of the SQP ratio for 1y ($\alpha= 99 \%$)}  \\	\cline{2-13}\\[-1.5ex]
$p=0$   & 0.36 & 0.55 & 0.39 & 0.50 & 0.51 & 0.60 & 0.57 & 0.70 & 0.51 & 0.47 & 0.51  & \bf{0.52 (0.09)} \\ 
$p=0.5$ & 0.45 & 0.48 & 0.40 & 0.46 & 0.43 & 0.59 & 0.49 & 0.64 & 0.47 & 0.45 & 0.54 & \bf{0.49 (0.07)} \\
$p=1$   & 0.45 & 0.48 & 0.40 & 0.44 & 0.42 & 0.54 & 0.46 & 0.59 & 0.46 & 0.42 & 0.52& \bf{0.47 (0.06)} \\
$p=2$   &  0.45 & 0.48 & 0.40 & 0.44 & 0.42 & 0.54 & 0.46 & 0.59 & 0.46 & 0.42 & 0.52& \bf{0.47 (0.06)} \\[2ex]
\hline
\end{tabular*}
\end{center}
\end{table}

As before, when considering the average SQP Ratio, we observe a similar - although slightly different - behavior for the average of all indices as with the S\&P 500 index (see the last two columns in Table~\ref{tbl-SQPratios-RMSE}). On average, the error increases with the threshold (except for $p=1$), and - as with the S\&P 500 - for $\alpha=95\%$ the error is smallest for $p=0.5$ and increases afterwards again. For $\alpha=99\%$ its maximum is reached for $p=0$, then the error levels off, being constant for $p=1,2$. Again, the average RMSE in all cases is clearly above 0 reaching 50\% of the right value 1 of the SQP ratio. This means that there are high fluctuations of the ratio, for the weak underestimation (of slightly less than 10\%) for $p=0$, as well as for a stronger overestimation for $p>0$ (recall the last column of  Table~\ref{tbl-avSQPR}). This analysis shows that the ratio alone cannot give a fine enough picture of the quality of risk estimation. Indeed, we want to understand why and in which circumstances we over- or underestimate the risk measured by the SQP throughout the sample. It is what we tackle in the next section.\\[-7ex]

%%%%%%%%%%%%%%%%%%%%%%%%%%%%%%%%%%%
\section{Analyzing and Quantifying pro-cyclicality}
\label{sec-ProcCycl}
\vspace{-2ex}
%%%%%%%%%%%%%%%%%%%%%%%%%%%%%%%%%%%

As mentioned in the introduction,  pro-cyclicality of risk measurements implies that, when the market is quiet, the future risk is under-estimated, while, during crisis, the future risk is over-estimated. In this section, we deal with this problem by first defining an indicator of market states, then  evaluating the performance of the risk measure when conditioning it on this indicator.}

\subsection{Volatility Impact on the Measured Risk}
\label{ssec-volImpact}
\vspace{-1ex}

\subsubsection{Realized Volatility as an Indicator of Market States}
\label{sssec-realVol}
\vspace{-2ex}
To better understand the dynamic behavior of the ratios, we condition them on past volatility. This is inspired by the empirical study on Foreign Exchange Rates done in \citet{Dacorogna2001}, where the authors conditioned the returns on past volatility. On the one hand, they showed that after a period of high volatility, the returns tend to be negatively correlated. It means that the volatility should diminish  in the future, with a distribution more concentrated around its mean; thus the quantiles should be smaller at given thresholds than current ones. It implies that the future risk would be overestimated when evaluated on this high volatility period. On the other hand, they showed that in times of low volatility, consecutive returns tend to be positively auto-correlated; thus the volatility should increase on the long term, with the consequence in the future of quantiles shifted to the right, meaning that our estimation would underestimate the future risk. 

To be able to condition on volatility, we define an empirically measured annualized volatility, $v_{k,T}(t)$, taken over a rolling sample of size $T$ year(s), corresponding to $n$ business days, up to but not including a time $t$. By abuse of notation, we refer with $t-j$ to a point in time $j$ days before time $t$ (for any integer-valued $j$), which gives us: 
\begin{eqnarray}\label{eq:annualVol}
& v_{k,T}(t)=\sqrt{n}\times v_{k,n}(t-1) \, , \quad \text{where} \quad  v_{k,n}(t-1) \;\text{denotes a realization of} \nonumber\\ [0.5ex]
& V_{k,n}(t-1) := \left \{ \frac{1}{n-1} \sum_{i=t-n}^{t-1} \Big | X_i-\frac{1}{n}\sum_{i=t-n}^{t-1} X_i \Big |^k \right \}^{1/k}, 
\end{eqnarray}
the $X_i$'s denoting the daily log-returns defined in \eqref{eq:ret}, and $k$ plays the same role as $p$ in~\eqref{eq:SQPDef_specified_mu}, emphasizing more or less large events. We consider $\displaystyle k=1,2$. If $k=2$, $v_{2,T}(t)$ is nothing else  than the usual empirical standard deviation (denoted std), called in the literature 'realized volatility', whereas if $k=1$, it is the empirical Mean Absolute Deviation (denoted MAD). We may generalize the notion of realized volatility to any $k$ and use it for $k=1,2$. 
Note that the notation is chosen such that the annual realized volatility $v_{k,T}(t)$ at time $t$ (January 1, 2014 for instance) is computed on the previous $T$ year(s) until $t-1$ included (i.e. until December 31, 2013, in our previous example). We evaluate $v_{k,T}(t)$ on a rolling basis every month. Hereby, we obtain a monthly time series of annual realized  volatility $(v_{k,T}(t))_t$, which can be used as a benchmark of the current market state.

\begin{figure}[h]
\begin{center}
  \includegraphics[width=12cm, height=6cm]{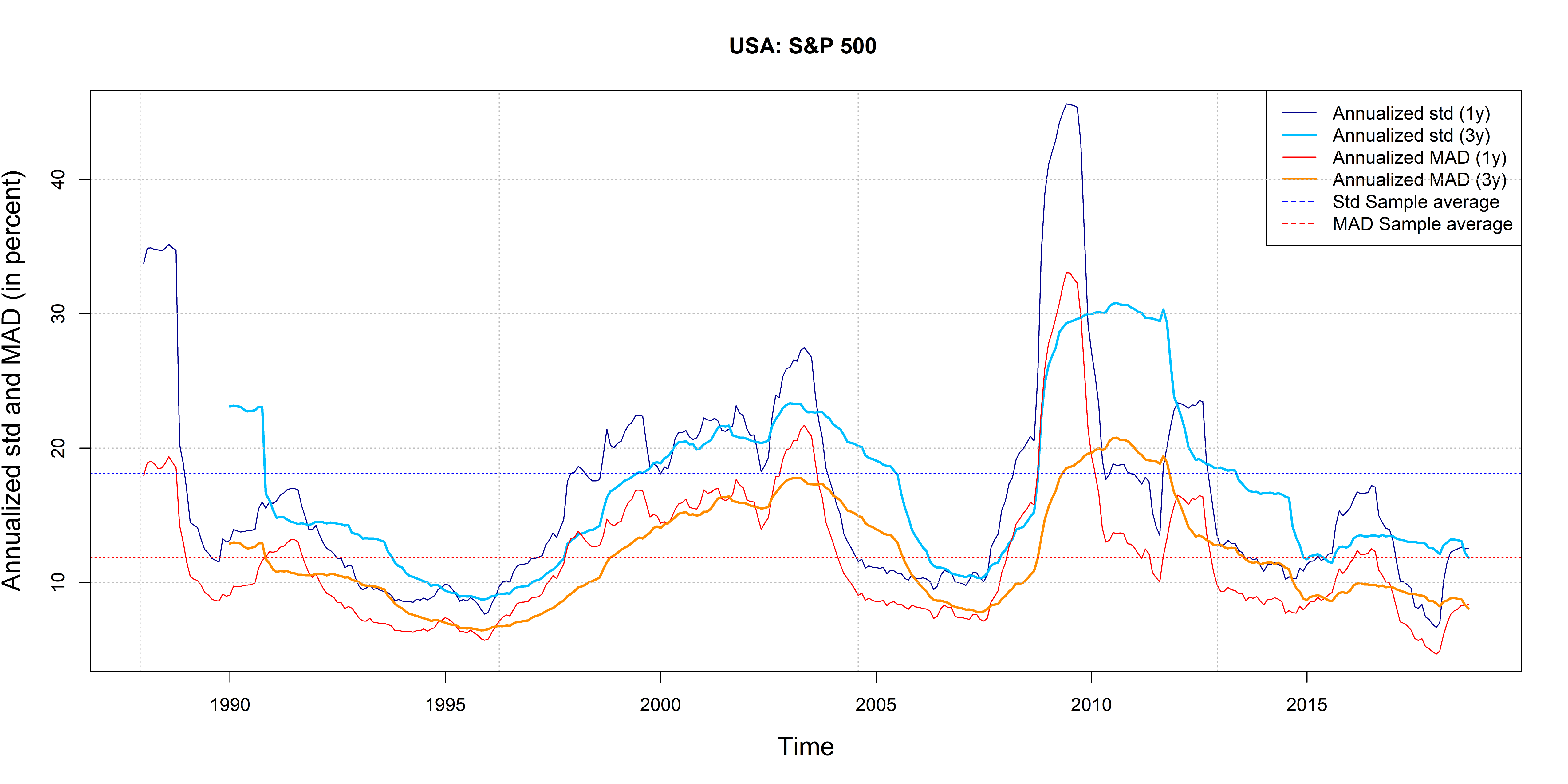}
\end{center}
  \caption{\label{fig:AnnVol}{\sf\small Annual realized volatility $v_{k,T}(t)$, defined in \eqref{eq:annualVol}, in percent for $k=1$ (MAD) and $k=2$ (std), of the S\&P 500 returns over a rolling sample of size $T=1,3$ year(s). The dashed lines represent the sample average  (over the whole sample) of the various measures of the volatility (with the same color code).}}
\end{figure}
In Figure~\ref{fig:AnnVol}, we present $v_{k,T}(t)$ for the S\&P 500 (log-)returns between January 1987 and September 2018 for both a rolling sample of $T=1$ year and of 3 years, as well as for $k=1$ and $k=2$. We also indicate the average annual realized volatility of the index calculated over the whole sample for both $k=1$ and $k=2$ (represented by the dashed lines). Their values are 11.9\% for $k=1$ and 18.1\% for $k=2$. We can see that the volatilities above these benchmarks mostly stand for periods of high instability or crisis (and not only in the USA). The high volatility of the period 1987-1989 is for instance explained by the New York Stock Exchange crash in October 1987. In 1997, Asia was hit by a crisis, as well as Russia in 1998 and Argentina in 1999-2000. In 2001, the United States experienced the bursting of the internet bubble. Following the Lehman Brother's bankruptcy, the period of 2008-2009 was a period of very high volatility. And finally, the sovereign debt crisis in Europe also impacted the S\&P 500 Index in 2011-2012; this is an illustration of the increased dependence between markets during times of crisis. Observing those co-movements in various indices during crises is important for portfolios of financial assets, when looking at pro-cyclicality of the global market.
 Moreover, we see that the dynamics of both definitions ($k=1$ and $k=2$) in \eqref{eq:annualVol} are quite similar, while the size of the movements are more pronounced in the case of the estimated standard deviation. 

We conclude from the analysis above that the annual realized volatility $v_{k,T}(t)$ for $k=1,2$ is a good indicator that reacts well and quickly to various market states. It is thus a reasonable proxy to qualify the times of high risks and to use it to discriminate between quiet periods and periods of crisis. We use it to condition our statistics (the SQP ratio \eqref{eq:VaRRatio}) and see if we can detect different behaviors of the price process (which underlies the risk measure) during these periods in comparison to quiet times, as it was the case for returns in~\citet{Dacorogna2001}. We turn now to this statistics to look at the ability of the SQP risk measures to correctly predict the future risks, whatever the market state, and to capture the dynamics.
Note that to ease the reading in the rest of the paper, we will use abusively in the text the term 'volatility' for the expression 'annualized realized volatility', $v_{k,T}$ (as well as for the rv $V_{k,T}$).

\subsubsection{SQP Predictive Power as a Function of Volatility}
\label{sssec-SQPandVol}
\vspace{-1ex}

\begin{figure}[h]
\vspace*{-0.5cm}
\hspace*{1.5cm}
\begin{minipage}{0.5\textwidth}
\includegraphics[scale=1.05,width=6cm,height=6cm]{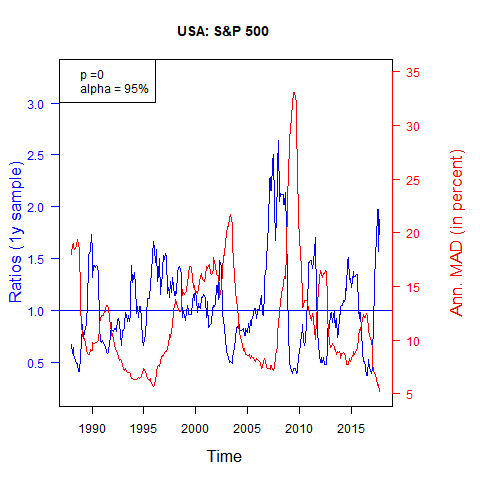}
\end{minipage}
\hspace{-1.5cm}
\begin{minipage}{0.5\textwidth}
\includegraphics[scale=1.05,width=6cm,height=6cm]{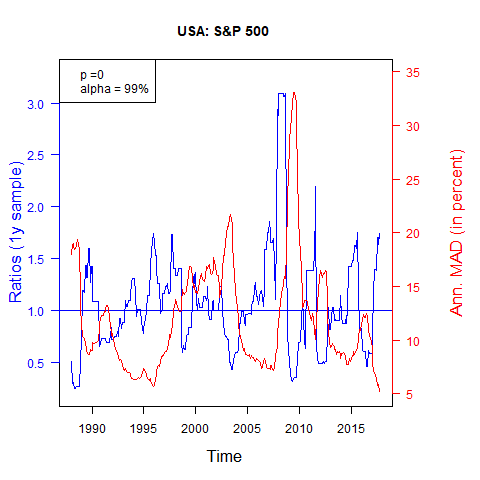}
\end{minipage}
\caption{\label{fig:RatioSQVol_p0} \sf\small Time series of SQP Ratios \eqref{eq:VaRRatio} for $p=0$ (left $y$-axis in dark) and of annualized realized volatility \eqref{eq:annualVol} for $k=1$ (MAD; right $y$-axis in light)  computed over 1 year ($T=1y$) for $\alpha = 95\%$ (left plot) and $\alpha = 99\%$ (right plot)}
\end{figure}

In the case of the S\&P 500, we represent in the same graph the time series of both the volatility $v_{k,T}(t)$ and the realized ratios of SQP's for both thresholds 95\% and 99\% to compare their behavior over time. It turns out that similar observations hold when using a 1 or 3-year sample to predict the estimated future risk and the volatility, as well as for the various $p$, and for $k=1,2$. 
% (see~\cite{Marcel2017} for more details in the case $k=2$). 
A further discussion of the influence of the sample size can be found in Section~\ref{sss-influences}. We illustrate in Figure~\ref{fig:RatioSQVol_p0} the dynamics  when considering e.g. the case of a 1-year sample, $p=0$ (rolling VaR) and $k=1$ for the volatility (MAD). One can easily observe on these graphs a strong opposite behavior for SQP ratios (in dark) and for volatility (in light). 

In Figure~\ref{fig:RatioSQVol_p0}, the dynamic behavior appears clearly. The realized ratios can differ from 1 either on the upper side or on the lower side. The negative dependence between the realized SQP ratios and the volatility is obvious on both graphs.
This particular feature is of importance in our study because it means that when the volatility in year $t$ is high in the market, the realized SQP ratio in year $t$ is quite low. Recalling the definition of the ratio, this situation means that the risks have been over-evaluated by the SQP calculated in year $t$ (in comparison to the realized risk in year $t+1y$). Conversely, when the volatility is low in year $t$, the realized SQP ratio is often much higher than 1 in this year, which means that the risks for the year $t+1$ have been under-evaluated by the calculation of the SQP in year $t$. 

\begin{figure}[h]
\hspace*{1.5cm}
\begin{minipage}{0.5\textwidth}
\includegraphics[width=6cm,height=6cm]{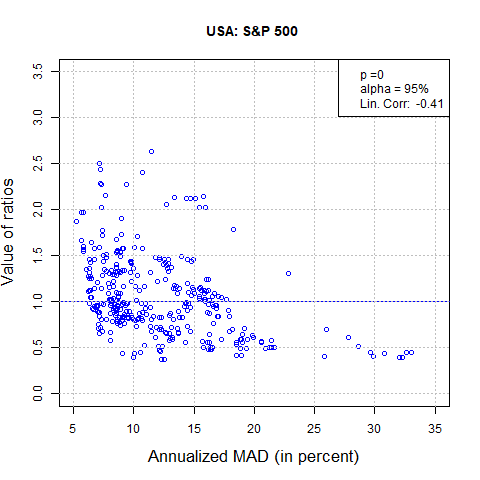}
\end{minipage}
\hspace{-1.5cm}
\begin{minipage}{0.5\textwidth}
\includegraphics[width=6cm,height=6cm]{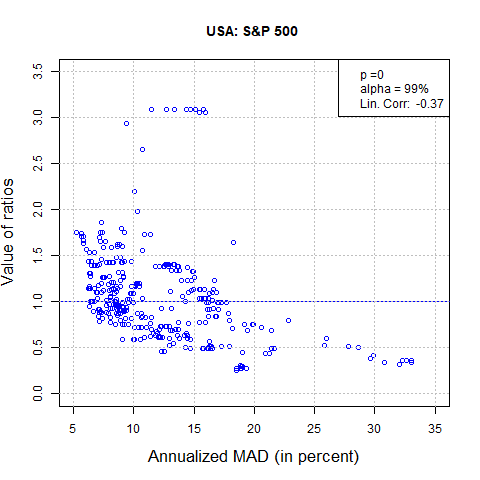}
\end{minipage}
\\ \hspace*{1.5cm}
\begin{minipage}{0.5\textwidth}
\includegraphics[width=6cm,height=6cm]{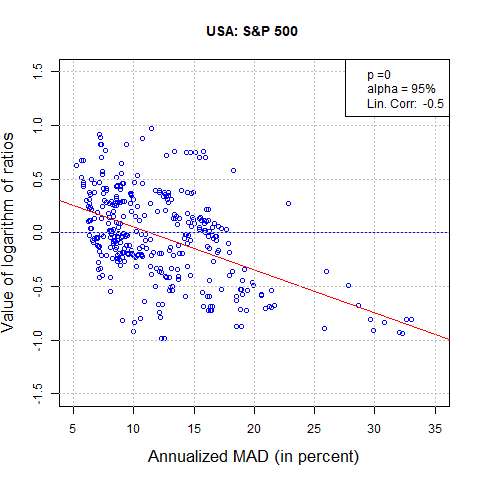}
\end{minipage}
\hspace{-1.5cm}
\begin{minipage}{0.5\textwidth}
\includegraphics[width=6cm,height=6cm]{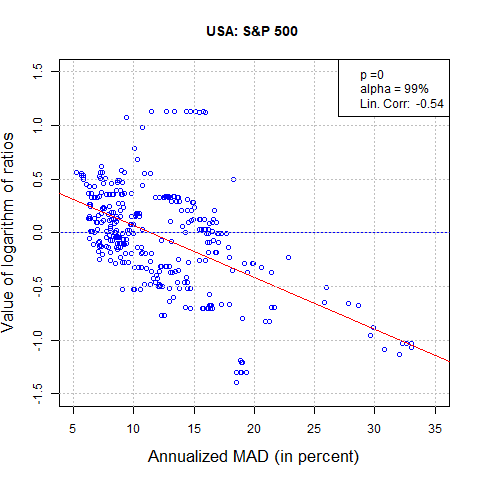}
\end{minipage}
\caption{\label{fig:SQRatio_p0} \sf\small SQP Ratios as a function of volatility (MAD) for $p=0$: On the left $\alpha = 95\%$ and on the right $\alpha = 99\%$;
first row SQP-ratios, second row logarithm of SQP-ratios}
\end{figure}

Another way of visualizing the pro-cyclicality is to plot the various realized SQP ratios as a function of volatility. We illustrate the obtained behavior in Figure~\ref{fig:SQRatio_p0}, choosing the same parameters as in Figure~\ref{fig:RatioSQVol_p0}.
Such figures highlight better the existing negative dependence between these two quantities. 
The plots of the first row of Figure~\ref{fig:SQRatio_p0} point out that the (negative) dependence is non linear, hence we consider the log-ratios instead (see the plots on the second row of Figure~\ref{fig:SQRatio_p0}), which provides a higher dependence than without the log: We observe that the more volatility there is in year $t$, the lower are the ratios $R_{p,\alpha,T}(t)$. When the log-ratios are negative, it means that losses in year $t+1$ have been overestimated with the measures calculated in year $t$. In the original scale, we can better quantify the magnitude of the effect: When looking at the case $\alpha=99$\%, with respect to the overestimation, we can observe ratios below 0.5 for high volatilities (above twice the average of 12\%, with MAD), {\it i.e.} the risk computed at the height of the crisis is more than twice the size of the risk measured a year later. On the other hand, we see that the underestimation can be very big since the realized ratios for the SQP can be even larger than 3; in other words, the risk next year is more than 3 times the risk measured during the current year for this sample. We also observe that the overestimation is very systematic for high volatility, whereas it is not for the underestimation at low volatility. Indeed, in this latter case, we see also values below 1, while we do not see any value above 1 when the volatility is high (above 20\%, close to twice the average over the whole sample). Although we present here only the S\&P 500, very similar behaviors are exhibited by all the other indices. % (see~\cite{Marcel2017}, again for $k=2$).

As already said, similar patterns can be observed for various choices of $p$, but it is of interest to find out for which $p$, the measure $\mu_p$ produces the strongest pro-cyclical behavior. To make this more quantifiable, given the highly non-linear behavior of this dependence (as illustrated in Figure~\ref{fig:SQRatio_p0}), we choose the {\it logarithm} of the SQP ratio (instead of the SQP ratio) to look at the linear Pearson correlation $\displaystyle \rho\left(\log R_{p,\alpha,T}(t), V_{k,T}(t)\right)$ (with $V_{k,T}(t)=\sqrt{n}\times V_{k,n}(t-1)$) considering the volatility choosing $k=1$ (MAD). We will keep this choice of volatility (MAD)  in the rest of the paper. Indeed, an important motivation, in particular for practical use, for this choice  is that it implies weaker conditions on the moments of the underlying distribution than when using the classical standard deviation ($k=2$): 
%, considered in \citet{Marcel2017}): 
For the existence of the correlation between the two quantities (the log ratio and the volatility respectively), we need the existence of the variances of those estimators, i.e. the existence of a second moment of the underlying distribution for MAD instead of a fourth one for the standard deviation. So, from now on, we use the name 'volatility' for 'annualized realized MAD' ($v_{1,T}$ abd the rv $V_{1,T}$). In Table~\ref{tbl-pcorr}, we present the value of $\rho(\log(R_{p,\alpha,T}(t), V_{k,T}(t))$ for all indices and as the average over all indices.

\begin{table}[h]
\begin{center}
\parbox{450pt}{\caption{\label{tbl-pcorr}\sf\small Pearson correlation $\rho(\log(R_{p,\alpha,T}(t), V_{k,T}(t))$ between the log of the SQP ratios and the volatility, for each index and for $T=1$ year, over the whole historical sample, and for two thresholds (95 and 99\%). In the last column, we present the average over all indices $\pm$ the standard deviation over the 11 displayed values.}}
\vspace{-2ex}
\scriptsize
\begin{tabular*}{450pt}{l c c c c c c c c c c c c}    \hline \\[-1ex]
 & AUS & CAN& FRA& DEU & ITA & JPN & NLD & SGP &SWE & GBR & USA & {\bf AVG ($\pm$ sd)} \\[1ex] \hline\hline
	\\[-1ex]
	& \multicolumn{11}{c}{$\alpha= 95 \%$}  \\	\cline{2-13}\\ [-1.5ex]
p=0   &  -0.51 & -0.46 & -0.50 & -0.49 & -0.52 & -0.58 & -0.52 & -0.50 & -0.51 & -0.50 & -0.50 & {\bf -0.51 $\pm$ 0.03} \\ 
p=0.5 & -0.57 & -0.43 & -0.46 & -0.43 & -0.43 & -0.59 & -0.53 & -0.52 & -0.43 & -0.53 & -0.47  & {\bf -0.49 $\pm$ 0.06} \\ 
p=1   & -0.48 & -0.39 & -0.37 & -0.18 & -0.24 & -0.35 & -0.51 & -0.50 & -0.27 & -0.48 & -0.45& {\bf -0.38 $\pm$ 0.11} \\ 
p=2   &  -0.50 & -0.38 & -0.27 & -0.11 & -0.21 & -0.33 & -0.35 & -0.45 & -0.15 & -0.40 & -0.35& {\bf -0.32 $\pm$ 0.12} \\[2ex]
 & \multicolumn{11}{c}{$\alpha= 99 \%$}  \\	\cline{2-13}\\[-1.5ex]
p=0   &   -0.57 & -0.46 & -0.51 & -0.47 & -0.52 & -0.52 & -0.55 & -0.54 & -0.48 & -0.50 & -0.54  & {\bf -0.51 $\pm$ 0.04} \\
p=0.5 & -0.49 & -0.41 & -0.36 & -0.21 & -0.32 & -0.36 & -0.41 & -0.46 & -0.27 & -0.43 & -0.42  & {\bf -0.38 $\pm$ 0.08} \\
p=1   & -0.49 & -0.43 & -0.37 & -0.20 & -0.36 & -0.37 & -0.41 & -0.47 & -0.30 & -0.41 & -0.43 & {\bf -0.39 $\pm$ 0.08} \\
p=2   &  -0.49 & -0.43 & -0.37 & -0.20 & -0.36 & -0.37 & -0.41 & -0.47 & -0.30 & -0.41 & -0.43& {\bf -0.39 $\pm$ 0.08} \\[2ex]
\hline
\end{tabular*}
\end{center}
\end{table}
We see that the correlation is always significantly negative, the average values for different thresholds and values of $p$ are above 30\% (in absolute value). The lower the value of $p$, the stronger the negative linear correlation.  
% In general one can say that the correlation is similar for both thresholds. 
The different $p$-values discriminate better the correlation behavior for 95\% than for 99\%.
One might also conclude that, for those observations, SQP with $p=0$ (rolling VaR) are better suited to emphasize the cyclical signal, than with higher values of $p$. 
% thus for designing counter-cyclical models.
 
Due to the non-linear nature of the dependence,  we also consider the notion of rank correlation, looking for instance at the Spearman $\rho_S$.\\[-5ex]
\begin{table}[H]
\small
\begin{center}
\parbox{340pt}{\caption{\label{tbl-Spearman}\sf\small Average over all indices of the Spearman $\rho_S$ between the volatility and the SQP ratios computed on a 1-year sample}}\\[-1ex]
\begin{tabular*}{210pt}{l c c}
\hline
\\[-1.5ex]
        &  $\alpha=95\%$ & $\alpha=99\%$  \\[0.5ex]
\hline\hline
\\[-1.5ex]
~~$p=0$   & {\bf -0.48 $\pm$ 0.05} & {\bf -0.49 $\pm$ 0.03} \\ 
~~$p=0.5$ & {\bf -0.45 $\pm$ 0.05} & {\bf -0.38 $\pm$ 0.05} \\
~~$p=1$   & {\bf -0.37 $\pm$ 0.08} & {\bf -0.39 $\pm$ 0.05} \\
~~$p=2$   & {\bf -0.31 $\pm$ 0.09} & {\bf -0.39 $\pm$ 0.05} \\  
[1ex]
\hline
\end{tabular*}
\end{center}
\end{table}
In Table~\ref{tbl-Spearman}, we display only the average Spearman $\rho_S$ over all indices as we have seen in the other tables that indices do not differ significantly from one another in their behaviors. 
% (the values for each index separately as well as for the Kendall tau are available in~\cite{Marcel2017}). 
In this table, the results confirm those obtained when considering the Pearson correlation between the logarithm of the SQP ratios and the volatility (e.g. highest value in absolute terms for low values of $p$ for both thresholds $\alpha$, decreasing in $p$ for $\alpha=95$\%, almost constant in $p$ for $p>0$ in the case of $\alpha=99$\%, when considering the volatility with $k=2$),
only the magnitude (and the standard deviation over the 11 indices) is slightly lower than for the Pearson correlation.

Note that we have performed the same analysis using a 3-year sample (T=3) to compute the estimated future risk $Q_{p,\alpha,T}(t)$ and the volatility, and then looking at the dependence between volatility and SQP ratios. On average, effects are similar in magnitude and with mostly similar tendencies for both Pearson and Spearman correlations. 
% The corresponding tables can be found in~\cite{Marcel2017} for $k=2$. 

Let us turn to the question of why we observe such (negative) dependence. It is the target of the next section, to look for factors that explain the pro-cyclical behavior observed so far.

\subsection{Looking for Explanations}
\label{ssec-expl}

We want to find out the reasons of the under-/overestimation of risk. To do so, we need to use models able to isolate effects. The first and simplest model we consider is iid rv's. 
The reason of such choice is to check if the very way we measure risk, independently of volatility models, creates some negative dependence between the log ratio and the realized volatility.

Then we explore if the volatility clustering present in the data is fast enough to produce pro-cyclicality.
To this aim, we choose the simple GARCH(1,1) model, to isolate the clustering effect by itself. Indeed, there are many GARCH type models in the literature (e.g. EGARCH, HARCH, HAR, etc...) that might reproduce the empirical regularities of the data better than this choice, but they would mix effects (e.g. EGARCH mixes clustering and skewness, HARCH and HAR mix various clustering at different frequencies); it would render the interpretation much more difficult: from which effect would the pro-cyclicality come?  We then calibrate the GARCH(1,1) to the data, considering first normal innovations, then Student ones to see if there would be any additional effect to the clustering when better calibrating the tails.

Since we observed that the choice of $p=0$ would emphasize the presence of pro-cyclicality, we focus on this case. But, the reasoning presented here clearly holds for $p>0$. Given the fact that GARCH is defined on $\sigma^2$, we will explore the two cases for eq.\eqref{eq:annualVol} of $k=1$ and $2$.

\subsubsection{Via IID Models}
\label{sss-iir}

Let us consider a rv $X$, with mean $\mu$, finite variance $\sigma^2$, distribution function $F_X$. We simulate 100'000 samples each of size $N$, $(X_1,\cdots, X_N)$, with parent rv $X$, from which we  estimate, on rolling sub-samples of size $n$, the statistics considered so far: the logarithm of the ratio \eqref{eq:VaRRatio} with $p=0$, and the volatility \eqref{eq:annualVol} for $k=1,2$. 
To compute those statistics, we produce time series of empirical estimators $(Q_{0,\alpha,T,t})_{t\ge 0}$, with $T=1$ year, corresponding to $n=252$ business days, denoted, for simplicity, by  $(Q_{\alpha,n}(t))_{t\ge 0}$ and $(V_{k,n}(t))_t$ computed on rolling samples. We estimate them. respectively, as $\displaystyle  Q_{\alpha,n}(t)= X_{( \lceil n\alpha \rceil ),t}$ where $\displaystyle \lceil x \rceil = \min\{ m \in \Z : m \geq x \}$ and $X_{(i),t}$ denotes the $i$th order statistics associated with the sample $(X_{t-i},\, i=1,\cdots,n)$, and $\displaystyle V_{k,n}(t) = \frac{1}{n-1}\sum_{i=1}^{n} |X_{t-i}-\bar{X}_{n,t}|^k$ where $\displaystyle \bar{X}_{n,t} = \frac{1}{n} \sum_{j=1}^{n} X_{t-j}$. We evaluate the  empirical Pearson and Spearman correlations between the time series of the logarithm of a ratio of sample quantiles, $\displaystyle \log\left(Q_{\alpha,n}(t+1y)/Q_{\alpha,n}(t)\right)$, and the time series of the sample volatility $\displaystyle V_{k,n}(t) $. Note that the sample quantiles $Q_{\alpha,n}(t)$ and $Q_{\alpha,n}(t+1y)$ are computed on subsequent but disjoint samples, i.e. the points in time are $t$ and $t+1y$ ($t$ plus one year), and the sample size $n$ is chosen as $n=252$ (which corresponds to 1 year of data as explained in Section~\ref{sssec-realVol}). Hence, those samples are disjoint by construction.

We consider two standard cases for $F_X$, the Gaussian and the Student (with various degrees of freedom) distributions, to study the effect of both light and heavy tail distributions. The results are displayed in Table \ref{tbl-iidcorr}, when taking $N\simeq 8000$ in the simulations  (to match the real data sample sizes), thresholds $\alpha=95$\%, $99$\% and $99.5$\%, and both definitions of volatility ($k=1,2$).
\begin{table}
\parbox{465pt}{\caption{\label{tbl-iidcorr}\sf \small Averages over 100000 repetitions of Pearson and Spearman correlation between $ \log\left(Q_{\alpha,n}(t+1y)/Q_{\alpha,n}(t)\right)$ and $V_{k,n}(t) $ for the IID case. Two asterisks indicate a negative value for the correlation at the 99\% confidence level, one at 95\%, none at the 90\%, and $\dagger$ at the 85\%. The values with no theoretical counterpart (because of infinite fourth moment) are put into brackets. On the right plot, comparison between $V_{2,n}(t)$ and $V_{1,n}(t)$ for the Pearson correlation. }}\\[-2ex]
\begin{minipage}{0.65\textwidth}
%\begin{center}
{\scriptsize
\begin{tabular}{l| cc | cc cc}
\hline
&&&&&&\\[-1ex]
  & \multicolumn{2}{c|}{Normal sample}  & \multicolumn{4}{c}{Student-t sample} \\
   & \multicolumn{2}{c|}{~~$X \sim \mathcal{N}(\mu, \sigma)$~~}  & \multicolumn{2}{c}{$X \sim t(5, \mu, \sigma)$} & \multicolumn{2}{c}{$X \sim t(3, \mu, \sigma)$}\\[1ex]
\hline\hline
&&&&&&\\[-1ex]
                &         k=1           &         k=2          &         k=1           &         k=2   &         k=1           &         k=2          \\%[1ex]
\hspace*{-5pt}{\bf {Pearson}} &&&&&&\\  
$\alpha=95\%$ & $-0.34^{\star \star}$ & $-0.40^{\star \star}$ & $-0.37^{\star \star}$ & $-0.31^{\star}$ & $-0.35^{\star \star}$ & $(-0.19)$ \\
$\alpha=99\%$ &        $-0.23$~~      & $-0.32^{\star \star}$ & $-0.29^{\star}$~      & $-0.36^{\star \star}$& $-0.33^{\star \star}$ & $(-0.29)$ \\
$\alpha=99.5\%$&       $-0.20^\dagger$      & $-0.29^{ \star}$ & $-0.27^{\star }$~      & $-0.37^{\star \star}$& $-0.32^{ \star}$ & $(-0.33)$ \\[1ex]
\hspace*{-5pt}{\bf {Spearman}} &&&&&&\\                         
$\alpha=95\%$ & $-0.33^{\star \star}$ & $-0.38^{\star \star}$ & $-0.35^{\star \star}$ & $-0.32^{\star \star}$ & $-0.34^{\star \star}$ & $-0.24$     \\
$\alpha=99\%$ &        $-0.22$~~      & $-0.31^{\star \star}$ & $-0.28^{\star}$~      & $-0.36^{\star \star}$ & $-0.32^{\star \star}$ & $-0.34^{\star \star}$ \\
$\alpha=99.5\%$ &        $-0.19^\dagger$      & $-0.28^{ \star}$ & $-0.26^{\star}$~      & $-0.36^{\star \star}$ & $-0.31^{\star \star}$ & $-0.37^{\star \star}$ \\[2ex]
\hline
\end{tabular}
}
%\end{center}
\end{minipage}
\begin{minipage}{0.35\textwidth}
  \includegraphics[scale=1.05,width=6cm, height=6cm]{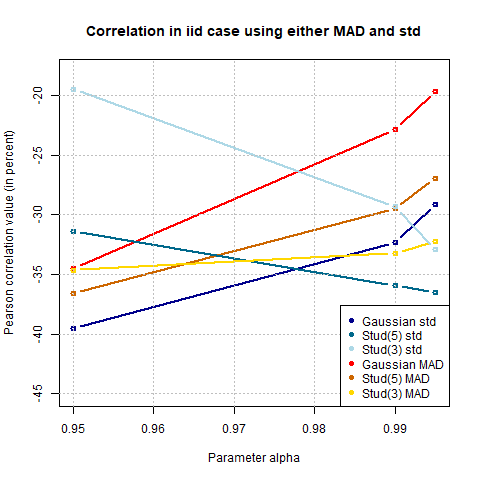}
  \end{minipage}
\end{table}
Note that the Gaussian distribution and the Student one with $\nu=5$ degrees of freedom have a finite fourth moment, hence we can use the Pearson correlation for both $k=1$ and $2$, whereas, for $\nu=3$, the theoretical counterpart value for $k=2$ does not exist, because of an infinite fourth moment (it means that the empirical values obtained in such a case have no statistical significance, even if computable because of a finite sample).
%; see \cite{Marcel2018}. 
We observe that, at high confidence levels and for any $k$, both Pearson and Spearman correlations are negative and of the same order. We notice also that those correlations are smaller in absolute value for $k=1$ than for $k=2$ when $\alpha=99$\% or $99.5$\%; for $\alpha=95$\%, the same comment holds true for the normal distribution, but it is the opposite for the Student distributions. 
% In addition, this is in line with the results obtained for all the financial indices when comparing with the study in~\citet{Marcel2017} for $k=2$. 
Finally, we also observe that the correlation decreases in absolute value when the threshold $\alpha$ increases, except for $k=2$ in the case of the Student distribution, as illustrated in the plot of Table ~\ref{tbl-iidcorr}. Those observations are in line with the theoretical results obtained in \cite{Marcel2018} (see their Fig.7, 3rd row, first two plots that correspond to $k=2$ and $k=1$ from left to right), where the impact of the volatility estimator choice on the rate of convergence of the correlation is pointed out.

Comparing the results in Table~\ref{tbl-iidcorr} with those obtained with historical data in Table~\ref{tbl-pcorr}, the correlation appears less pronounced in the former (for $\alpha=99\%$, the Pearson correlation is $-0.51$ for the average overall the stock indices, compared to $-0.29$ for Student-t ($\nu=5$) iid rv's). So, the pro-cyclicality is only partially explained by the way risk is measured. Some part of the negative correlation stays yet not explained. In the next section, we turn to the GARCH model to see if volatility clustering might explain the remaining effect.

\subsubsection{Via GARCH Modelling}
\label{sss-garch}

\paragraph{Fitting the GARCH -}

In our empirical analysis, we observe volatility clustering that goes, of course, hand in hand with a return to the mean of volatility. Thus, we think that part of the pro-cyclical behavior with volatility might be explained by this clustering, since the measuring is done on a longer time period than the typical clustering size. A natural model, namely the GARCH, comes to mind as it has been developed precisely for modelling volatility clustering. Recall that our goal is not to fit as accurately as possible the data (choosing the latest generation of GARCH type models) but to isolate causes for the pro-cyclical behavior.

Given this aim, we turn to  the simplest version of GARCH, the GARCH(1,1), to model the log- returns $(X_t)_t$ as: $\displaystyle \, X_{t+1} = \epsilon_t\,\sigma_t$, where the variance $\sigma_t^2$ of $X_t$ satisfies
\begin{equation}\label{eq:garch}
\sigma_t^2 = \omega + \alpha \, X_t^2 + \beta \, \sigma_{t-1}^2, \quad \text{with}\quad\omega>0, \,\alpha>0, \, \beta>0, 
\end{equation}
where the error terms $(\epsilon_t)_t$ constitute either a Gaussian or a Student-t white noise (with mean 0 and variance 1).
Note that, in the context of GARCH models, we always refer to the standard deviation when using the term `volatility', i.e. $v_{k,T}$ with $k=2$.

To estimate our model, we fit the parameters $\omega, \alpha, \beta$ to each full sample of the 11 indices, using a robust optimization method proposed by Zumbach (see in\citet{Zumbach2000}), which is the most precise in reproducing the average volatility compared to the other methods. As a side note, we remark that the robustness is especially important as other optimization methods may not always fulfill the stationarity condition ($\alpha + \beta < 1$) for the GARCH process.

Even though we consider a simple model, some care is needed to find a good way to optimize the GARCH(1,1) parameters with the right choice of innovations, in order to reproduce both the clustering effect and the accurate tail behavior present in the data. To do so, we argue on both the maximum likelihood and the $\tau_\text{cor}$ criteria, where $\tau_\text{cor}$ is a parameter specifying the time decay $\delta t$ of the autocorrelation of the square log returns, expressed in terms of $(\alpha,\beta)$: $\tau_\text{cor}:=\delta t/|\log(\alpha+\beta)|$  (see Eq.(12) in \citet{Zumbach2000}). Let us explain how we arrive at the optimal model: We start with \eqref{eq:garch} optimized with, respectively, Gaussian and Student innovations on the full samples. We observe that the model with Gaussian innovations provides a good fit for the $\tau_\text{cor}$ and the mean volatility, whereas the tail is underestimated. Switching from Gaussian to Student innovations increases the likelihood, which is good, but also the $\tau_\text{cor}$, worsening the fit of the autocorrelation of the square log returns. Moreover, the tail is still underestimated particularily for the negative returns (losses). 
Therefore, we fit the GARCH with Student innovations on the losses only, obtaining reasonable tails, with degrees of freedom $\nu$ around 5-6, but obviously too short $\tau_\text{cor}$ and smaller normalized likelihood. Hence, to optimize both the $\tau_\text{cor}$ and the tail, we choose the model combining the GARCH (1,1) parameters obtained with Gaussian innovations on full samples, with Student innovations fitted on the losses only. We check that this model has a better likelihood than the one with Gaussian innovations, and a similar one as for the model optimized on full samples with Student innovations. Moreover, this model combines a good fit for the $\tau_\text{cor}$ and the mean volatility, and a good tail estimation. 
\begin{table}[h]
\caption{\label{tbl-garch-param}\sf\small Fitting results of the GARCH models, defined in \eqref{eq:garch}, for the 11 indices}
\scriptsize
\begin{tabular}{c ccc ccc ccc cc}
\hline
\\[-1.5ex]
        & AUS & CAN & FRA & DEU & ITA & JPN & NLD & SGP & SWE & GBR & USA  \\
\hline\hline
\\[-1.5ex]
\hspace*{-15pt}{\bf { Gaussian innovations}} \\
$\omega$ [$10^{-6}$] & 3.45 & 1.11 & 2.54 & 2.18 & 2.42 & 5.29 & 2.21 & 2.78 & 3.29 & 1.81 & 1.70 \\ 
$\alpha$ [$10^{-1}$] & 1.58 & 1.03 & 0.93 & 0.87 & 0.80 & 1.42 & 1.07 & 1.23 & 1.05 & 1.06 & 0.99 \\
$\beta$ [$10^{-1}$] & 8.06 & 8.85 & 8.90 & 8.99 & 9.06 & 8.27 & 8.81 & 8.60 & 8.75 & 8.77 & 8.88 \\[0.5ex]
$\alpha+\beta$          & {\it 0.96} & {\it 0.99} & {\it 0.98} & {\it 0.99} & {\it 0.99} & {\it 0.97} & {\it 0.99} & {\it 0.98} & {\it 0.98} & {\it 0.98} & {\it 0.99}
\\ $\tau_{cor}$ (days) & 28& 86& 60&  71&  70&  32&  80&  60&  51&  58&  76 
\\[1ex] 
\hspace*{-5pt}{\it Fitting Results}\\
Normalized Likelihood & 3.37 &3.45 &3.12& 3.15& 3.05& 3.04& 3.13& 3.16& 3.09 &3.34& 3.27 \\ 
Volatility [\%]     & 15.7 &15.5 &19.5 &19.7 &20.6 &20.9 &20.9 &20.4 &20.6 &16.3 &17.9  \\
Historical [\%]     & 15.7 &15.5& 19.6& 19.7& 20.7& 21.0& 21.1& 20.6& 20.7& 16.4& 18.1
\\[1ex] 
\hspace*{-15pt}{\bf Student innovations (fitted }\\
\hspace*{-15pt}{\bf on losses) with }\\
\hspace*{-15pt}{\bf Gaussian GARCH parameters}\\
$\nu$      & 5.15 & 4.75 & 5.97 & 6.00 & 5.98 & 5.89 & 5.77 & 5.09 & 6.44 & 6.25 & 4.75   
\\[1ex] 
\hspace*{-5pt}{\it Fitting Results}\\
Normalized Likelihood &  3.40 & 3.47 & 3.14 & 3.17 & 3.07 & 3.07 & 3.15 & 3.20 & 3.12 & 3.36 & 3.30 \\ 
Volatility [\%]     & 15.4 &14.9& 19.5& 19.5& 20.5& 20.8& 20.6& 19.7& 20.6& 16.2 &17.5  \\
\hline
\end{tabular}
\end{table}

\begin{figure}[H]
\centering
\begin{minipage}{0.6\textwidth}
\includegraphics[width=7cm,height=7cm]{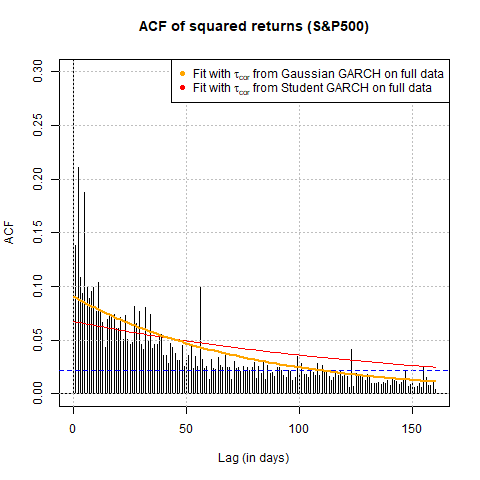}
\end{minipage}
\hspace{-2.5cm}
\begin{minipage}{0.4\textwidth}
\parbox{200pt}{\caption{\label{fig:ACF-tauCor} \sf\small Autocorrelation (ACF) of the squared returns of the S\&P500 with the fitted $\tau_\text{cor}$ for Gaussian (yellow) and, respectively, Student (red) innovations. The (blue) dashed line represents the 95\% confidence level for the ACF.}}
\end{minipage}
\end{figure}

In Table~\ref{tbl-garch-param}, we report the GARCH parameters, as well as the results for $\tau_\text{cor}$ and the normalized likelihood (for fair comparison between the samples of different sizes) - for both, the GARCH model with Gaussian innovations as well as the optimized model keeping the Gaussian GARCH parameters $(\omega,\alpha,\beta)$, but considering now  Student innovations fitted on the losses only. We see that the optimization gives, in all cases, parameters that fulfill the GARCH stationarity condition. The estimated parameters do not vary much from one index to the other, except for AUS and JPN. Those two indices exhibit the shortest $\tau_\text{cor}$. The typical clustering obtained with this fit is between 5 weeks (shortest period for AUS) and 4 months, with an average value of 3 months (a business month contains 21 days), which is short enough to produce pro-cyclicality on a yearly horizon.
As can be seen in Figure~\ref{fig:ACF-tauCor}, the $\tau_\text{cor}$ for the Gaussian innovations reproduces better the clustering of the data than for the Student ones, which present a too long extent.
\vspace{-1ex}
\vspace{-3ex}

Although we do not expect a large variation for the likelihood, as it is not so sensitive to tails, we observe in Table~\ref{tbl-garch-param} a slight increase of the normalized likelihood for all indices,  when going from Gaussian to Student innovations; it indicates a consistent improvement in the description of the data (when fitting other models already discussed above, this slight increase did not show up for all indices as here). 

As an additional check for the goodness of fit, we report in Table~\ref{tbl-garch-param}, for each index, the average volatility obtained by simulation of the fitted GARCH (with 1000 replications for each index, considering the same sample size as the original data sample), and compare it with the average annualized historical volatility over the real data. We see that the volatilities of the GARCH reproduce very well (slightly lower) the historical ones on average, when estimated over 1000 replications of the GARCH process, whatever the innovation, although slightly better for the Gaussian innovations. 

Then, considering both, the GARCH model with Gaussian innovations as well as the optimized one with Student innovations (but keeping the Gaussian GARCH parameters), we apply the same statistical analyses as those done on the historical data in Section~\ref{ssec-volImpact} on each of the 1000 replications in both cases. 
This means we analyze the behavior of the SQP for varying parameters $p$ and thresholds $\alpha$, we look at the RMSE and the SQP-ratios; as those results are not central to our analysis, we do not present them here. Then we look closer at the Pearson correlation between volatility and the logarithm of the SQP-ratios; it is what is developed next. All in all, we observe qualitatively similar results with the GARCH simulations as with the historical data and the values are closer to those of the indices than in the case of iid rv's.\\[-7ex]

\paragraph{Dependence between SQP Ratios and Annualized Realized Volatility}
\label{par-dpdGarch}
~

\begin{figure}[h]
\hspace*{1.5cm}
\begin{minipage}{0.5\textwidth}
\includegraphics[width=6cm,height=6cm]{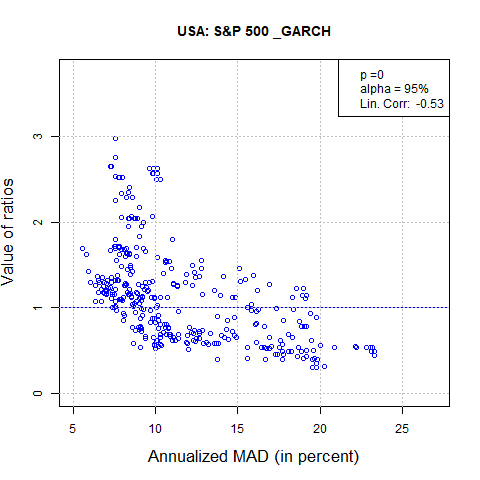}
\end{minipage}
\hspace{-1.5cm}
\begin{minipage}{0.5\textwidth}
\includegraphics[width=6cm,height=6cm]{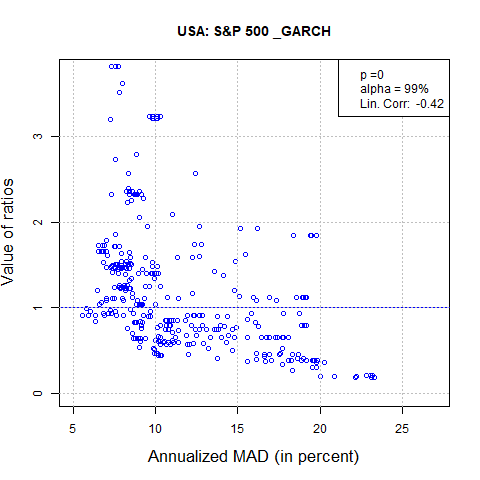}
\end{minipage}
\\ \hspace*{1.5cm}
\begin{minipage}{0.5\textwidth}
\includegraphics[width=6cm,height=6cm]{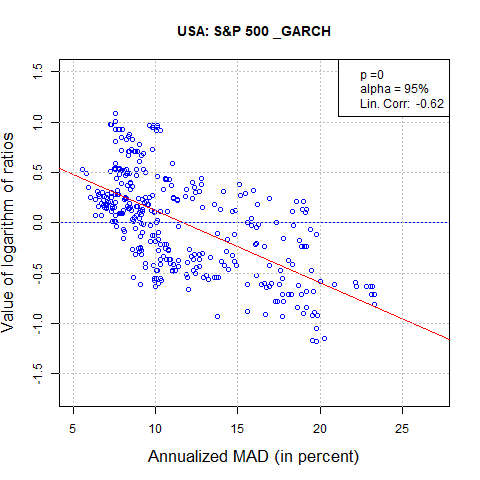}
\end{minipage}
\hspace{-1.5cm}
\begin{minipage}{0.5\textwidth}
\includegraphics[width=6cm,height=6cm]{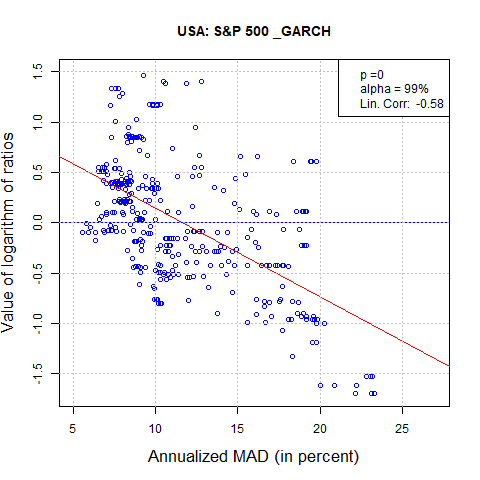}
\end{minipage}
\caption{\label{fig:SQRatio_GARCH_avg} \sf\small SQP Ratios as a function of volatility (MAD) for one simulated sample path of the GARCH(1,1) (that fits the S\&P500) for $p=0$: On the left $\alpha = 95\%$ and on the right $\alpha = 99\%$;
first row SQP-ratios, second row logarithm of SQP-ratios}
\end{figure}

As in Section~\ref{sssec-SQPandVol}, we first look at the Pearson correlation between the annualized daily volatility (recall, we focus on the MAD) and the logarithm of the SQP ratios computed on a 1-year sample. Those are depicted in Table \ref{tbl-corr_SQP_annVol_Garch_avg}: We observe negative correlations of the GARCH for the averages, as for historical data. Yet, they are higher (in absolute value) than for the data (see Table \ref{tbl-pcorr}), the difference between the two increasing with $p$. Furthermore, as in the case of historical data, the negative correlation is stronger for the lower powers $p$, but is systematically decreasing (in absolute value) with the threshold, which is not the case for the data. Note that the choice of innovations has a weak impact on the correlation values, almost vanishing with higher thresholds. The parameter $p$ does not discriminate any longer with increasing thresholds (e.g. for 99.5\%, the correlations are identical for $p>0$). 
\begin{table}[H]
%\begin{center}
\parbox{465pt}{\caption{\label{tbl-corr_SQP_annVol_Garch_avg}\sf\small Average (over all indices and its 1000 repetitions each) Pearson correlation between volatility and log of SQP ratios computed on a 1-year sample, using a GARCH(1,1) model, and for three thresholds (95, 99, and 99.5\%). Standard deviation is computed over the 11 averages of the values obtained for each index.}}\\[-1ex]
\scriptsize
\begin{tabular*}{460pt}{l cc cc cc}
\hline
\\[-1.5ex]
~~AVG$\pm \sigma$  & \multicolumn{2}{c}{$\alpha=95\%$} & \multicolumn{2}{c}{$\alpha=99\%$} & \multicolumn{2}{c}{$\alpha=99.5\%$} \\
                   & ~~~Gaussian & Student-t & ~~~Gaussian & Student-t & ~~~Gaussian & Student-t  \\[0.5ex]
\hline\hline
\\[-1.5ex]
~~$p=0$   & ~~~{\bf -0.63 $\pm$ 0.01} & {\bf -0.63 $\pm$ 0.01} & ~~~{\bf -0.58 $\pm$ 0.01 } & {\bf -0.59 $\pm$ 0.01} & ~~~{\bf -0.57 $\pm$ 0.01 } & {\bf -0.57 $\pm$ 0.01} \\ 
~~$p=0.5$ & ~~~{\bf -0.64 $\pm$ 0.02} & {\bf -0.66 $\pm$ 0.01} & ~~~{\bf -0.57 $\pm$ 0.01} & {\bf -0.57 $\pm$ 0.01} & ~~~{\bf -0.55 $\pm$ 0.01 } & {\bf -0.56 $\pm$ 0.01} \\
~~$p=1$   & ~~~{\bf -0.63 $\pm$ 0.02} & {\bf -0.65 $\pm$ 0.01} & ~~~{\bf -0.56 $\pm$ 0.01} & {\bf -0.57 $\pm$ 0.01} & ~~~{\bf -0.55 $\pm$ 0.01 } & {\bf -0.56 $\pm$ 0.01} \\
~~$p=2$   & ~~~{\bf -0.59 $\pm$ 0.01} & {\bf -0.61 $\pm$ 0.01} & ~~~{\bf -0.56 $\pm$ 0.01} & {\bf -0.57 $\pm$ 0.01} & ~~~{\bf -0.55 $\pm$ 0.01 } & {\bf -0.56 $\pm$ 0.01} \\
[1ex]
\hline
\end{tabular*}
%\end{center}
\end{table}
\vspace{-2ex}
As in Figure~\ref{fig:SQRatio_p0}, we illustrate the obtained results with the S\&P 500 index when considering one (simulated) sample path of the fitted GARCH(1,1) model (the one with Gaussian GARCH parameters but with Student innovations). The simulated sample path depicted was chosen such that it is representative of the average behavior observed in the simulation. We look at the various realized SQP ratios (as well as its logarithm) for $p=0$,  as a function of the volatility; see Figure~\ref{fig:SQRatio_GARCH_avg}. For a better comparison between the GARCH realization and the historical data, we use the same y-scale for both. %The experiment has been replicated with many sample paths and gives the same conclusion.

We observe a similar behavior for GARCH (Fig.\ref{fig:SQRatio_GARCH_avg}) and historical data (Fig.\ref{fig:SQRatio_p0}):
\begin{itemize}
\item
The correlation is negative in both cases and the slope of the linear regression lines look similar. 
\item\vspace{-.5ex}
The more volatility there is in year $t$, the lower than 1 are the ratios $R_{p,\alpha,T} (t)$, which means that losses in year $t+1$ have been overestimated with the measures calculated in year $t$.
\item\vspace{-.5ex}
This overestimation can result in ratios below 0.5: The risk computed at the height of the crisis, in the above realization, is more than double the size of the risk measured a year later. 
\item\vspace{-.5ex}
For the underestimation, we see that the realized ratios can be larger than 3; in other words, the risk next year is more than three times the risk measured during the current year for this sample. 
\item\vspace{-.5ex}
As with the historical data, this overestimation is systematic for high volatility, whereas it is less the case for low volatility.
\end{itemize}

We notice in Figure~\ref{fig:SQRatio_GARCH_avg}, that the linear correlation is not an optimal approximation of the dependence structure (as already observed for historical data, although the correlation is slightly stronger for GARCH). Thus, we consider the logarithm of the SQP ratio to take into account the non linear dependence and, additionally, compute a rank correlation (Spearman)- see Table~\ref{tbl-Spearman-rho_Garch_avg}.
\begin{table}[H]
\begin{center}
\parbox{460pt}{\caption{\label{tbl-Spearman-rho_Garch_avg}\sf\small Average (over all indices and its 1000 repetitions each) Spearman's $\rho_S$ rank correlation between the volatility and the SQP ratios computed on a 1-year sample, using a GARCH(1,1) model. Standard deviation is computed over the 11 averages of the values obtained for each index.}}\\[-1ex]
\scriptsize
\begin{tabular*}{460pt}{l cc cc cc}
\hline
\\[-1.5ex]
~~AVG$\pm \sigma$  & \multicolumn{2}{c}{$\alpha=95\%$} & \multicolumn{2}{c}{$\alpha=99\%$} & \multicolumn{2}{c}{$\alpha=0.995$} \\
                   & ~~~Gaussian & Student-t & ~~~Gaussian  & Student-t & ~~~Gaussian & Student-t \\[0.5ex]
\hline\hline
\\[-1.5ex]
~~$p=0$   & ~~~{\bf -0.61 $\pm$ 0.01} & {\bf -0.61 $\pm$ 0.00} & ~~~{\bf -0.56 $\pm$ 0.00} & {\bf -0.57 $\pm$ 0.01} & ~~~{\bf -0.55 $\pm$ 0.00} & {\bf -0.56 $\pm$ 0.01} \\ 
~~$p=0.5$ & ~~~{\bf -0.61 $\pm$ 0.01} & {\bf -0.64 $\pm$ 0.01} & ~~~{\bf -0.55 $\pm$ 0.01} & {\bf -0.55 $\pm$ 0.01} & ~~~{\bf -0.53 $\pm$ 0.01} & {\bf -0.54 $\pm$ 0.01} \\
~~$p=1$   & ~~~{\bf -0.61 $\pm$ 0.01} & {\bf -0.63 $\pm$ 0.01} & ~~~{\bf -0.55 $\pm$ 0.01} & {\bf -0.56 $\pm$ 0.01} & ~~~{\bf -0.53 $\pm$ 0.01} & {\bf -0.54 $\pm$ 0.01} \\
~~$p=2$   & ~~~{\bf -0.57 $\pm$ 0.01} & {\bf -0.59 $\pm$ 0.01} & ~~~{\bf -0.55 $\pm$ 0.01} & {\bf -0.56 $\pm$ 0.01} & ~~~{\bf -0.53 $\pm$ 0.01} & {\bf -0.54 $\pm$ 0.01} \\[1ex]
\hline
\end{tabular*}
\end{center}
\end{table}
For the GARCH(1,1) model, the rank correlation is of the same order (slightly lower) than the Pearson correlation. This is similar to the historical data (there the differences were a bit more pronounced, especially for low $\alpha$ and low $p$). On average (i.e. when looking at the average values obtained for the 11 indices considered) we see, in this case, that the behavior remains similar with respect to $\alpha$ and $p$ for the Pearson and Spearman correlation.

Overall, a simple GARCH model captures pretty well the main features we have seen in the data, with some differences discussed above. We conclude that the clustering of volatility and its return to the mean appear as an important factor of the pro-cyclicality, reinforcing the effect of the risk measurement itself, discussed in the iid case.

%%%%%%%%%%%%%%%%%%%%%%%%%%%%%%%%%%%%%%%%%
\subsubsection{Other Influences}
\label{sss-influences}
%%%%%%%%%%%%%%%%%
Let us eventually discuss three other possible influences on pro-cyclicality: sample sizes, data frequency and type of risk measures. 

% Remarks on the influence of sample sizes and data frequency
First, we look at the impact of sample sizes. Recall that we estimate the correlation $\rho(\log R_{p,\alpha,T} (t), V_{k,T}(t))$, which we can generalize for different sample sizes $T_1,T_2,T_3$ to
\[ \rho \left (\log \frac{Q_{0,\alpha,T_1} (t+T_1 y)}{Q_{p,\alpha,T_3} (t)}, V_{k,T_2}(t) \right ). \]
Nevertheless, the setting of our study gives us natural restrictions on the choice of these three sample sizes, which we discuss in the following. In the study, $T_1 = 1$ was fixed as 1 year. This is the time frame prescribed by regulation in order to assess the risk of the next year. Hence, increasing or decreasing this sample size $T_1$ does not make sense when wanting to assess the pro-cyclicality in regulatory risk measurement. The volatility estimator $V_{k,T_2} (t)$ acts as a proxy for the market state. We have seen, in Figure~\ref{fig:AnnVol}, that choosing $T_2 = 1,3$ (years) work well for this. However, choosing longer sample sizes than 3 years to evaluate the volatility does not answer any longer our problem as we would not capture anymore the dynamics of the market state: Indeed, the longer the sample size $T_2$, the more $V_{k,T_2}$ will converge to the overall sample average killing the dynamics. As a third component, we can vary the sample size $T_3$ in the estimate $Q_{p,\alpha, T_3} (t)$ for the future risk. Again, when increasing $T_3$ too much will loose the dynamics and with it the main goal of correctly estimating $Q_{p,\alpha, 1} (t+1y)$. Also, note that choosing different sample sizes $T_2 \neq T_3$ would deprive the interpretability of our analysis: The volatility estimator at time $t$ should be computed over the same sample as the estimate for the future risk, otherwise it makes no sense to set these two quantities into relation with each another. Therefore, it makes sense to keep $T_1 =1$ and choose a reasonable and identical sample size for the two other quantities under study, $T_2,T_3$, no longer than 3 years.

The different claims can be confirmed empirically. E.g. when running the method for 5, 7 or 10 years we see that the longer the sample size, the faster disappears the dynamics as well as the clustering. Comparing the case of the 1 year sample size with the 3 years sample size, 
% (see e.g. the complete analysis obtained in the case of $T=3$ for $V_{2,T}$ in \cite{Marcel2017}), 
we still observe similar patterns in the analysis for the historical data in the case of  1 or 3 years. However, in the GARCH case, we can clearly observe the effects of the clustering: As expected, the longer the sample, the less visible the clustering, which means a lower dependency of the ratio with volatility than in the case of  $T=1$ year. Note also that, the higher the threshold, the longer the sample needs to be to see this effect of decreasing correlation.

A similar statement can be made for the iid case, confirmed by the theoretical study by \cite{Marcel2018} where a factor $1/\sqrt{1+T}$ appears in the asymptotic correlation when considering $T$ years for both the estimated SQP and the volatility (see Table 7 for location-scale models and Theorem 10 in the general case). It means that the shorter the sample size, the stronger the pro-cyclical impact of the risk measurement. For the GARCH case, this statement needs to be qualified: It is true as long as the sample size is bigger than the size of the volatility clustering. If the size is shorter than the clustering, we expect that the pro-cyclical effect of GARCH will be less apparent as it is mainly due to the return to the mean of the volatility.

The second influence we measure is the impact of the data frequency on our statistics. We run the same analysis on weekly data and obtain results for the negative correlation of the same order. For the iid case, only the number of points matters, which implies that the longer the frequency, the larger the time window must be to keep the same number of points for estimation. 

The third influence we analyze is the definition of risk measure. When turning from VaR to another popular risk measure such as Expected Shortfall (ES), we believe that similar statements could be made, judging from the iid case. Indeed, in \cite{Marcel2018}, results are available for general functionals of VaR and made explicit for ES, showing a similar behavior for VaR and ES (see e.g. Figure 4 in this quoted paper) with underlying iid samples. Moreover, we have seen above that the  volatility clustering reinforces the negative correlation between the SQP ratio and the volatility: There is no reason why this would reverse in the case of ES given the fact that ES can be seen as a weighted average of VaR's (see \citet{Ekt2015}).\\[-7ex]

%%%%%%%%%%%%%%%%%%%%%%%
\section{Conclusion}
\label{sec-conc}
\vspace{-1ex}
%%%%%%%%%%%%%%%%%%%%%%%

In this study, we explore the appropriateness of measuring risk with regulatory risk measures on historical data to forecast the risk faced by financial institutions. 

First, we check and quantify empirically, considering 11 stock indices, that risk measures based on VaR are in fact pro-cyclical: In times of low volatility, the risk is underestimated, while in times of crises the risk is overestimated. Although pro-cyclicality of risk models has widely been assumed by market actors, it was never clearly quantified, nor shown empirically with enough evidence. 
The identification and quantification of pro-cyclicality is made possible by modelling the risk measure as a stochastic process and conditioning it on volatility to be able to discriminate the different market states. For this, we use the Sample Quantile Process (SQP), a 'dynamic generalization of VaR'. To measure its predictive power, we introduce a new statistics, a look-forward ratio of risk measures. While we only used it for the VaR/SQP, it may be applied to any other risk measure like Expected Shortfall without restrictions. It is a novel way of measuring the accuracy of risk estimation that does not rely on number of violations.

Second, we look for factors explaining this pro-cyclical behavior. To do so, we introduce simple models to isolate effects at the source of pro-cyclicality. We find firstly that measuring risk on past historical data with a quantile, or a quantiles function, is intrinsically pro-cyclical, since even iid random variables present a negative dependence between the look-forward SQP ratio and the volatility. Moreover, the clustering of volatility modelled by a GARCH(1,1) process with Gaussian or Student innovations reinforces this negative dependence, as the volatility will tend to return to its mean in a much shorter period than a one year horizon ($\tau_{cor}$ is typically of only a few weeks).
We thus relate secondly and partly this pro-cyclical effect to the volatility clustering present in financial markets.

This work is one more step in the constant search for improving risk management and the way company assess risk in the future. Traditional risk measures have been well discussed and analyzed in a static approach. Since an important issue, unsolved yet, is the pro-cyclical effect of measuring risk on historical data, it is natural to move to dynamic risk measures to analyze this effect. In the introduction section, we have reviewed various approaches to study this effect and remedy it with macro-prudential rules. In this study, we follow another route: We first aim at quantifying such effect and determine its main intrinsic factors. It then makes it possible, in the future, to design new dynamic risk measures, simple enough to be of practical use, but also able to properly respond to the various market states in order to avoid pro-cyclicality.

To conclude, let us recall that the traditional risk measures, advocated in the current regulation, favor pro-cyclical behaviors of the market actors. Following a pure historical estimation of risk, leaves financial institutions unprepared in case of crisis. Consequently, regulation should, in fact, enhance the capital requirements in quiet times and relax them during the crises, i. e. one needs, as regulators are aware, to introduce anti-cyclical risk management rules. However, these rules should be designed without hampering the system or weakening the principles of economic valuation of liabilities, as it is currently done. Having quantified a pro-cyclical effect of the sample VaR and identified some of its causes, should help build an anti-cyclical risk measure. This is what we are currently developing: A design of a SQP with the proper dynamical behavior as a good basis for improving forecast of risk, in view of the next generation of regulation.\\[-8ex]

\section*{Acknowledgments} 
\vspace{-4ex}

We are grateful to Prof. Jiro Akahori for pointing out the SQP and for fruitful discussions on it. We also thank Dr. Gilles Zumbach for making us aware of his empirical studies, specifically on iid rv's, and interesting exchanges. Dr. Michel Dacorogna would like to acknowledge the support of LabEx MME-DII when holding the international chair LabEx MME-DII - ESSEC CREAR during Spring 2018.

\bibliographystyle{plainnat}
\bibliography{Lit}

\newpage

%reset the figure and table counter to appendix labeling:
\setcounter{figure}{0} \renewcommand{\thefigure}{A.\arabic{figure}} 
\setcounter{table}{0} \renewcommand{\thetable}{A.\arabic{table}}

\section*{APPENDIX}
\vspace{-1ex}

\begin{appendices}

In this appendix we elaborate on three supplementary analyses.
The first, in Appendix~\ref{app:resid}, tries to give further evidence why we can conclude that the pro-cyclicality observed in the data is partially due to the GARCH effects, i.e. return-to-the-mean and clustering of the volatility.

The two others in Appendices~\ref{app:regr} and~\ref{app:vol_bins} offer alternative ways of looking at the pro-cyclicality analysis of Section~\ref{ssec-volImpact}. Either by measuring the pro-cyclicality via linear regression (see~\ref{app:regr}) or interpreting it more in line with the risk management point of view (see~\ref{app:vol_bins})

\section{Pro-cyclicality Analysis of Residuals} \label{app:resid}

To strengthen our claim that the pro-cyclicality in the real data is (apart from its intrinsic part) due to the return-to-the-mean and the clustering of the volatility, i.e. the characteristic properties of a simple GARCH(1,1) model, we proceed as follows (for each index).
Recalling our GARCH(1,1) model
\begin{align*}
X_{t+1} &= \epsilon_t\,\sigma_t, 
\\ \text{with~}\sigma_t^2 &= \omega + \alpha \, X_t^2 + \beta \sigma_{t-1}^2
\end{align*}
we compute the empirical residuals $\hat{\epsilon}_t := X_{t+1} / \hat{\sigma}_t$ where we initialize $\hat{\sigma}_t$ by using one year of data (as burn-in sample).
As GARCH parameters, we use the ones fitted in Section~\ref{sss-garch}.
Then, we we compute the sample correlation between the sample quantile and the sample MAD on the time-series of residuals $\hat{\epsilon}_t$ (and not the real data itself!). This time series of residuals should be iid distributed with mean zero and variance 1.
The asymptotic bivariate normality of the sample quantile with either the MAD or the standard deviation for iid samples was proven in \cite{Marcel2018} and explicit formulas for the Gaussian and Student distribution are presented therein too. With these results we can compute the asymptotic correlation of the SQP-logratio and the MAD for iid samples (see \cite{Marcel2018}).

Recalling that we are computing a sample correlation, we use the theoretical asymptotic value of the correlation to provide the corresponding confidence intervals for the sample Pearson linear correlation coefficient (using the classical variance-stabilizing Fisher transform of the correlation coefficient for a bivariate normal distribution to compute the confidence intervals - see the original paper \cite{Fisher21} or e.g. a standard encyclopedia entry \cite{Rodriguez82}).
%original paper on p217 (section 3 p215-218) % or p1377 in the encyclopedia in the article about correlation)
Note that those confidence interval values have to be considered with care. The bivariate normality of the sample quantile estimator and sample MAD holds asymptotically. Hence, it is not clear if for the sample sizes considered, we can assume bivariate normality (this could be tested). Still, we provide those theoretical confidence intervals as approximate guidance.
We then verify if the sample correlation of the residuals falls in these confidence intervals, and how the sample correlation on the real data behaves in comparison.

\begin{figure}[H]
\centering

\begin{tabular}{ccc}
%\hline
\subf{\includegraphics[width=49.4mm]{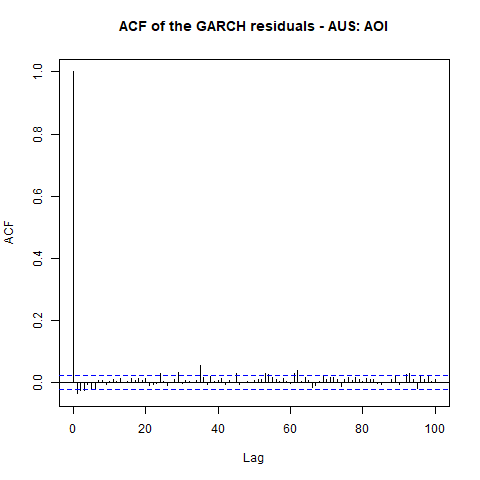}}
     {}
&
\subf{\includegraphics[width=49.4mm]{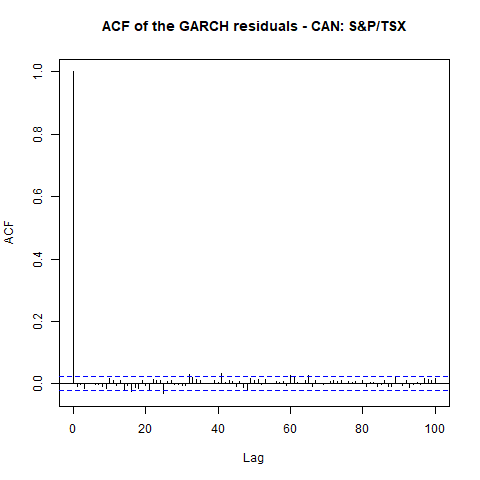}}
     {}
&
\subf{\includegraphics[width=49.4mm]{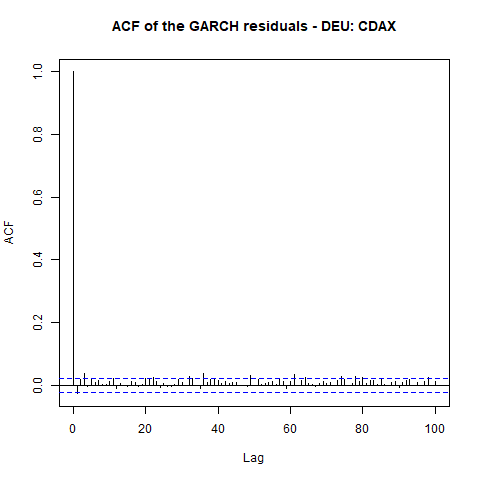}}
     {}
\\
%\hline
\subf{\includegraphics[width=49.4mm]{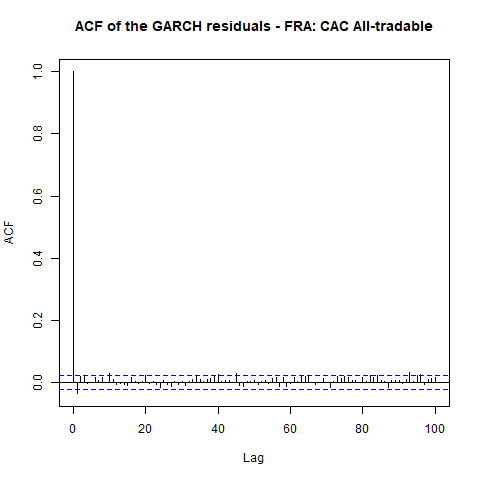}}
     {}
&
\subf{\includegraphics[width=49.4mm]{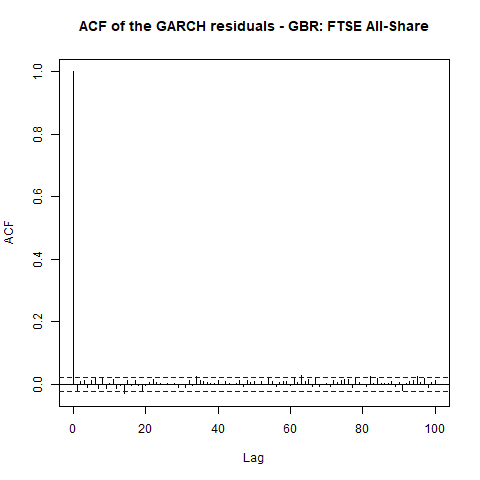}}
     {}
&
\subf{\includegraphics[width=49.4mm]{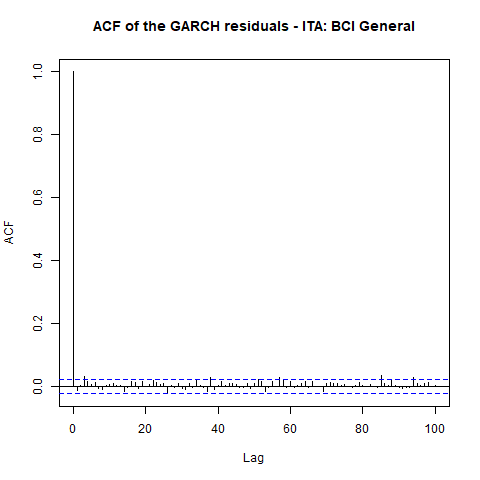}}
     {}
\\
%\hline
\subf{\includegraphics[width=49.4mm]{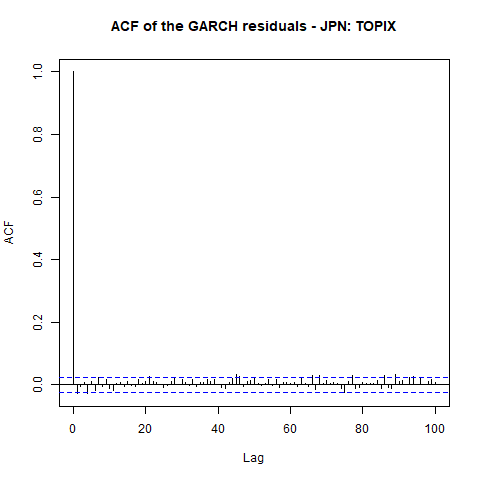}}
     {}
&
\subf{\includegraphics[width=49.4mm]{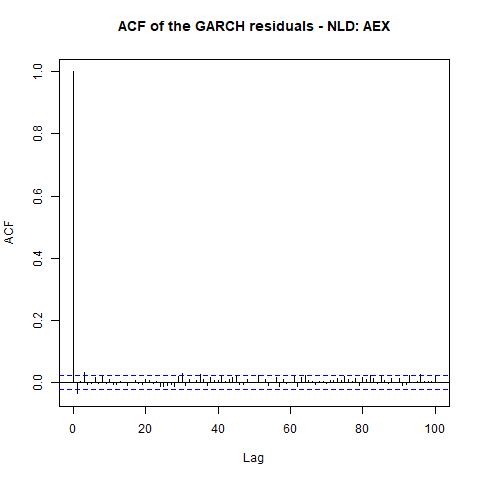}}
     {}
&
\subf{\includegraphics[width=49.4mm]{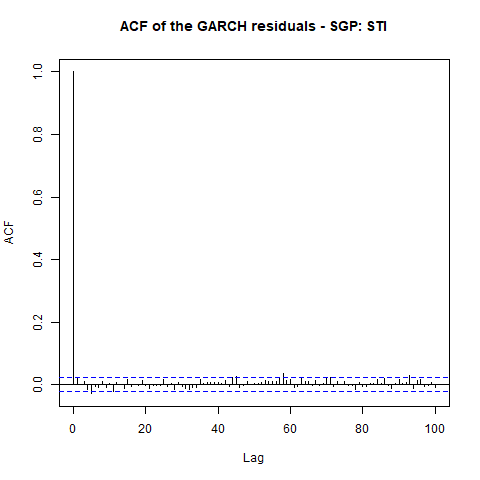}}
     {}
\\
%\hline
\subf{\includegraphics[width=49.4mm]{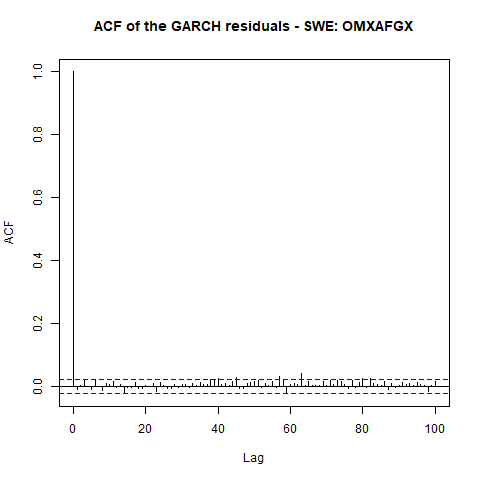}}
     {}
&
\subf{\includegraphics[width=49.4mm]{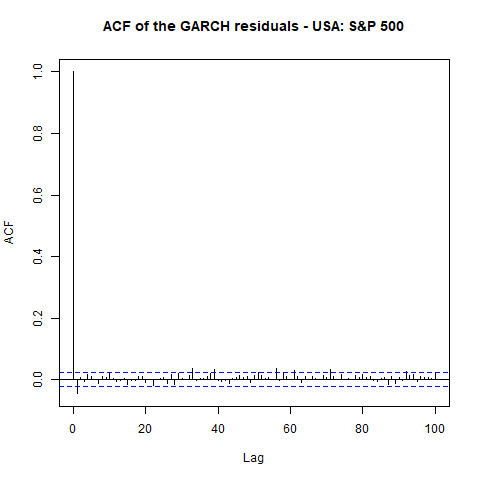}}
     {}
&
\\
%\hline
\end{tabular}
%\vspace*{-5mm}
\caption{\label{fig:ACF_residuals}\sf\small Plots of the autocorrelation function of the GARCH residuals for each index separately.}
\end{figure}

Before considering the plots of the correlation of the residuals, we first assess the mean and standard deviation in Table~\ref{tbl-residual-stats} and the autocorrelation plots of the absolute values in Figure~\ref{fig:ACF_residuals}, to see if the iid-assumption can be considered as reasonable.
Looking at the values in the table, we observe that they are close to the theoretically expected mean $0$ and standard deviation $1$.
Also, the autocorrelation plots show almost no autocorrelation within the first 100 lags. Thus, we can proceed with our analysis. 
\begin{table}[H]
\begin{center}
\parbox{470pt}{\caption{\label{tbl-residual-stats}\sf\small Mean and standard deviation for the empirical residuals of the GARCH(1,1) fits for each index. In the last column, we present the average over all indices $\pm$ the standard deviation over the 11 displayed values.}\vspace{-1ex}}
\footnotesize
\begin{tabular*}{480pt}{l c c c c c c c c c c c c}    \hline \\[-1ex]
% Mean for 
~~ & AUS & CAN& FRA& DEU & ITA & JPN & NLD & SGP &SWE & GBR & USA & {\bf AVG ($\pm~\sigma$)} \\[1ex] \hline\hline
	\\
% [-1ex]
%	& \multicolumn{11}{c}{SQP ratio for 1y ($\alpha= 95 \%$)}  \\	\cline{2-13}\\ [-1.5ex]
~mean    &   0.02  &0.03 & 0.02&  0.03&  0.01& -0.01&  0.03&  0.02&  0.03 & 0.02  &0.03 & {\bf 0.02} $\pm$ 0.01 \\ 
[1ex]
~Std. dev.    &1.00 &1.00 &1.00 &0.99 &1.00 &1.01 &0.99& 1.00& 1.00& 0.99& 1.00 & {\bf 1.00} $\pm$ 0.01\\
[2ex]
\hline
\end{tabular*}
\end{center}
\vspace{-2ex}
\end{table}

\newgeometry{top=0.75in, bottom=1.5in, left=1in, right=1in}
%\begin{adjustwidth}{}{-0.8cm}
%\end{adjustwidth}

\newgeometry{top=0.75in, bottom=1.5in, left=1in, right=1in}
%Different options with regard to CI and burn-in period
%90CI, 0.5y: 9out out of 44(0 times data in resid CI)
%95 CI, 0.5y: 7out of 44(0 times data in resid CI)
%99 CI, 0.5y: 1 out of 44 (but also 2times data in resid CI)

%90CI, 1y: 9out out of 44(0 times data in resid CI)
%95 CI, 1y: 6out of 44(0 times data in resid CI)
%99 CI, 1y: 2 out of 44 (but also 2times data in resid CI)
\vspace*{-2cm}
\begin{figure}[H]
\centering
%\begin{adjustwidth}{}{+2.6cm}
\begin{tabular}{ccc}
%\hline
\subf{\includegraphics[width=49.4mm]{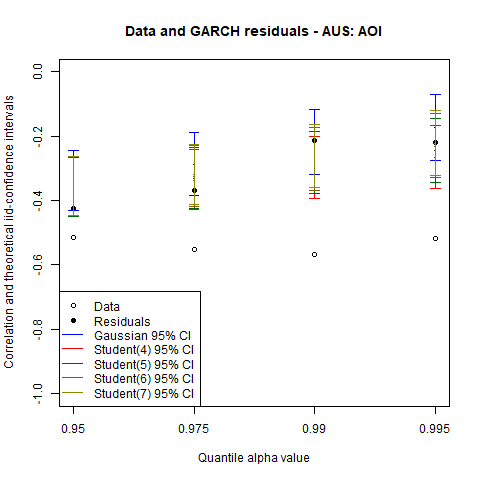}}
     {}
&
\subf{\includegraphics[width=49.4mm]{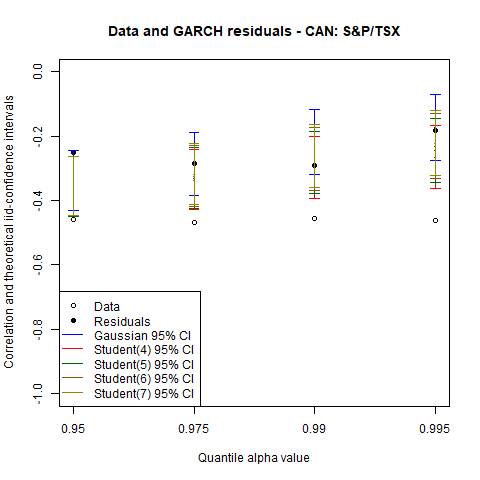}}
     {}
&
\subf{\includegraphics[width=49.4mm]{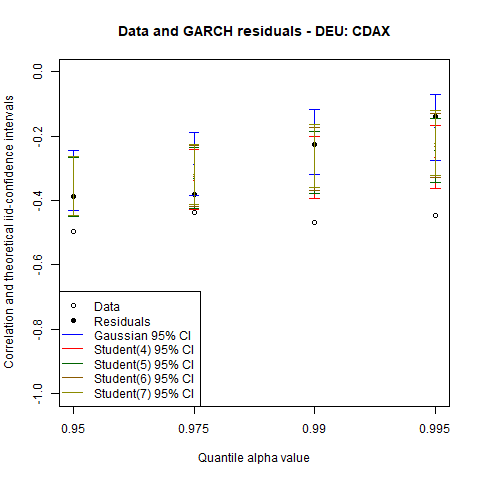}}
     {}
\\
%\hline
\subf{\includegraphics[width=49.4mm]{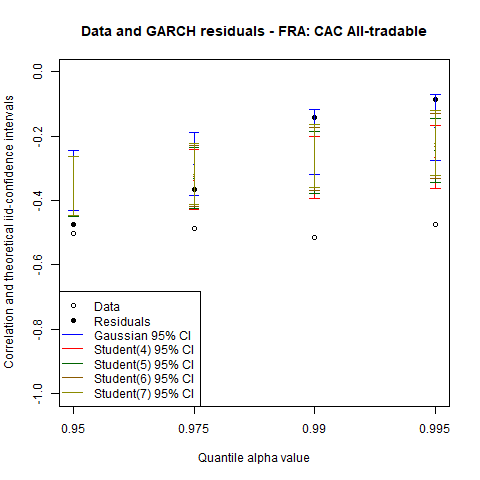}}
     {}
&
\subf{\includegraphics[width=49.4mm]{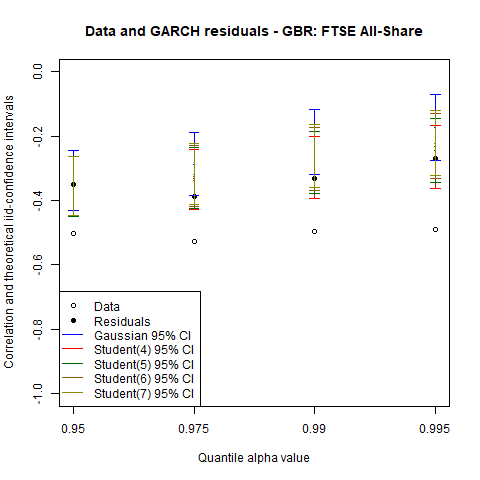}}
     {}
&
\subf{\includegraphics[width=49.4mm]{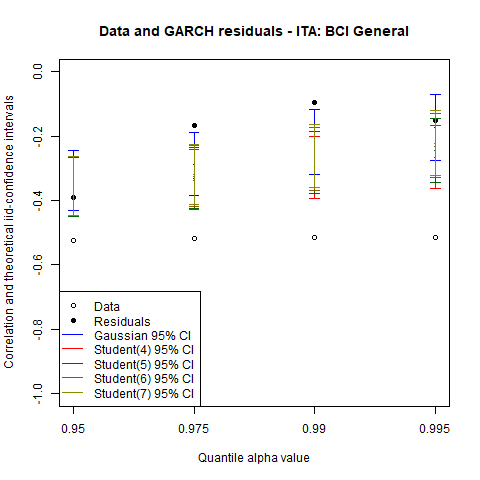}}
     {}
\\
%\hline
\subf{\includegraphics[width=49.4mm]{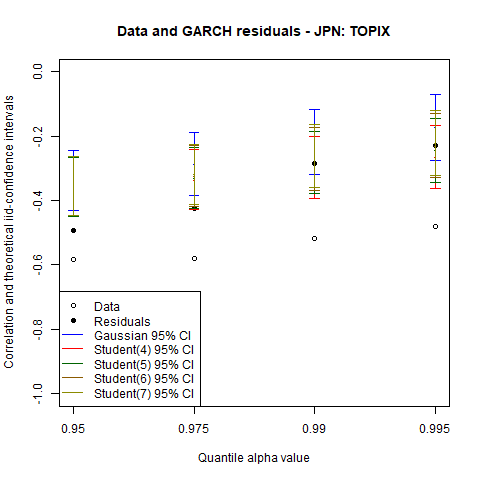}}
     {}
&
\subf{\includegraphics[width=49.4mm]{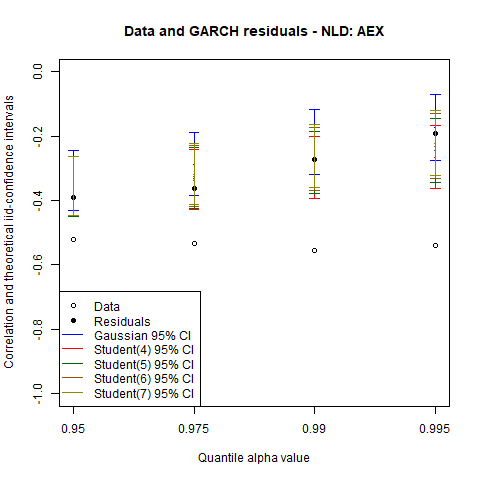}}
     {}
&
\subf{\includegraphics[width=49.4mm]{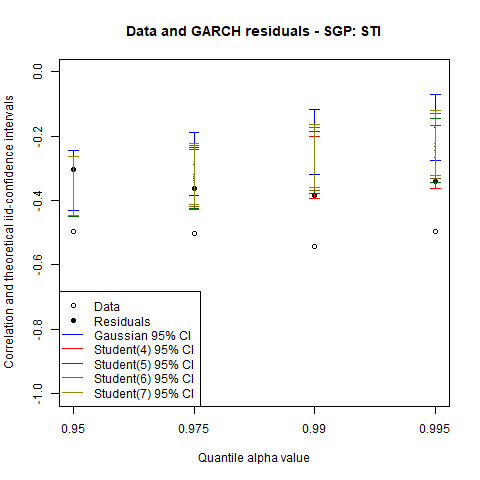}}
     {}
\\
%\hline
\subf{\includegraphics[width=49.4mm]{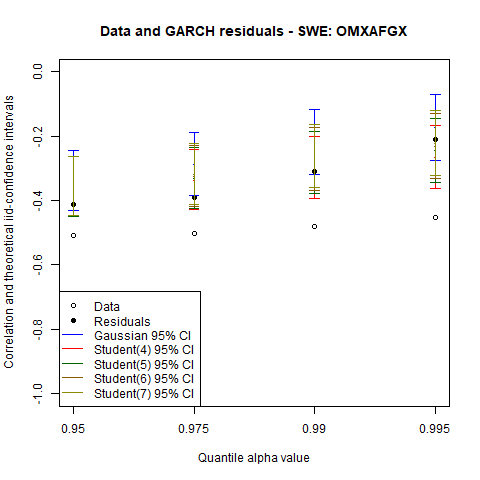}}
     {}
&
\subf{\includegraphics[width=49.4mm]{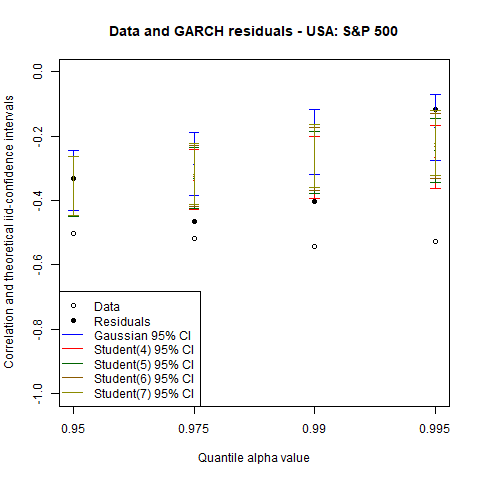}}
     {}
&
\\
%\hline
\end{tabular}
%\end{adjustwidth}
\caption{\label{fig:cor_residuals_vs_iid}\sf\small Comparison of correlation in the data with the correlation of the GARCH residuals for each index separately.
Each plot contains the correlation for the four different $\alpha$ values. For each of them, corresponding theoretical confidence intervals (for the sample correlation) assuming a specific underlying distribution (Gaussian or Student with different degrees of freedom) are plotted.}
\end{figure}

\restoregeometry
In Figure~\ref{fig:cor_residuals_vs_iid} we have one plot for each index. In each plot, we compare for each threshold $\alpha$ ($0.95,0.975,0.99,0.995$) the sample correlation (between sample quantile and sample MAD) on the real data versus the one on the residuals.
Further, 95\%-confidence intervals for a sample correlation assuming an underlying iid distribution are given - considering as alternatives a Gaussian or Student distribution, the latter with varying degrees of freedom, $\nu =4,...,7$.
In 38 out of 44 cases (86\%), the sample correlation of the residuals falls in the 95\% confidence interval of the sample correlation of an iid distribution. But in none of the cases the sample correlation of the real data falls in these confidence intervals.
Thus, we can claim that the correlation behaviour of the residuals seems to be as the correlation behaviour of iid random variables.
This finally means that stripping-off the GARCH features of the data by considering its residuals, we are left with a pro-cyclicality behaviour like for iid data. Hence, our claim that the correlation in the real data is due to two factors. One, the inherent pro-cyclicality due to using historical estimation, and second due to the GARCH effects, namely return-to-the-mean and clustering of volatility.

\section{Pro-cyclicality Analysis via Regression} \label{app:regr}

Recall, that in Section~\ref{sssec-SQPandVol} we computed the Pearson correlation between the (logarithm of the) look-forward ratio and the sample MAD, i.e. $\rho\left(\log R_{p,\alpha,T}(t), V_{1,T}(t)\right)$. Instead, to measure the linear dependence of the log-ratio with the volatility, we could have also used one of the following simple or multiple linear regression models: 
\begin{align} 
\log\lvert \frac{q_{n,t+1y}}{q_{n,t}} \rvert& = \alpha + \gamma \hat{\theta}_{n,t} + \epsilon_t,  \label{eq:regr}
\\ \log\lvert q_{n,t+1y} \rvert& = \alpha + \beta \log \lvert q_{n,t} \rvert + \gamma \hat{\theta}_{n,t} + \epsilon_t, \label{eq:mult_regr}
\end{align}
where the standard errors $(\epsilon_t)$ are assumed to be iid normally distributed with mean zero and constant variance.

First note that those hypotheses on the residuals are quite strong, and violated empirically here. 
While under such violated conditions the precision of the parameter estimation deteriorates, the results can still be used. But, we cannot take into account the standard errors or p-values and consequently cannot formally do a hypothesis test on the estimates.
It is the reason why we prefered proceeding via Monte-Carlo simulations on various models to explore the significance of the negative correlation between the look-forward ratio and the volatility (estimated via the MAD). 

Second note that the multiple linear regression model, \eqref{eq:mult_regr}, while appealing for its structure (all quantities depending on time $t$ are on one side of the equation), does not model the question we are trying to answer. We are not claiming to be able to predict the future value $q_{n,t+1y}$ from the knowledge of $q_{n,t}$ and $\hat{\theta}_{n,t}$! In out work we only set into relation the look-forward {\it ratio} of $q_{n,t+1y}$ and $q_{n,t}$ (i.e. a non-linear function of $q_{n,t}$ and $q_{n,t+1y}$) with  $\hat{\theta}_{n,t}$. Thus, the simple linear regression model is the apropriate one for us.
%Testing this correlation on many data sets (simulated from different models) confirms the validity of the negative values observed empirically on all the real data sets. Moreover, this approach allows us to identify possible causes for such effect.

Going on and testing this simple linear regression model, \eqref{eq:regr}, we obtain, on all stock indices, a negative $\gamma$ estimate with a small $p$-value (under the violated condition of iid residuals with constant variance), see Table~\ref{tbl-regr}. 
This means that we would accept the null hypothesis of $\gamma<0$ which corresponds exactly to antithetic behavior of look-forward ratio and volatility as described in Section~~\ref{sssec-SQPandVol}.

\begin{table}[H]
\begin{center}
\scriptsize
\parbox{440pt}{\caption{\label{tbl-regr}\sf\small Results on the regression estimate of $\gamma$ in~\eqref{eq:regr}, additionally standard error of $\gamma$ and $p$-value of the hypothesis test $\gamma <0$ - for each of the 11 stock indices and two thresholds (95 and 99\%). In the last column, we present the average over all indices $\pm$ the standard deviation over the 11 displayed values.}\vspace{-2ex}}
\begin{tabular*}{440pt}{l c c c c c c c c c c c c}    \hline \\[-1ex]
 & AUS & CAN& FRA& DEU & ITA & JPN & NLD & SGP &SWE & GBR & USA & {\bf AVG ($\pm~\sigma$)} \\[1ex] \hline\hline
	\\[-1ex]
	& \multicolumn{12}{c}{Regression estimates for $\alpha= 95 \%$}  \\	\cline{2-13}\\ [-1.5ex]
estimate & -4.9 & -3.9 &-4.5 &-4.1& -4.8& -5.9& -4.1& -3.8& -4.5& -5.0& -4.0 & -4.5 (0.6)
\\ std. error &  0.43 &0.40&0.41& 0.38 &0.42& 0.44& 0.36 &0.35&0.40 &0.46 &0.36 & 0.40 (0.04)
\\ $p$-value &  0 & 0 &0 &0& 0& 0 &0 &0 &0 &0& 0& 0 (0)
\\ & \multicolumn{12}{c}{Regression estimates for $\alpha= 99 \%$}  \\	\cline{2-13}\\ [-1.5ex]
estimate & -6.1 &-4.7 &-4.2 &-4.1& -4.9& -6.4& -4.6& -5.9& -4.5& -5.8& -4.8 & -5.1 (0.8)
\\ std. error&  0.47 &0.48 &0.37 &0.41& 0.43& 0.56& 0.37& 0.49& 0.44& 0.53& 0.40 & 0.45 (0.06)
\\ $p$-value &  0 & 0 &0 &0& 0& 0 &0 &0 &0 &0& 0& 0 (0)
 \\[2ex]
\hline
\end{tabular*}
\end{center}
\vspace{-2ex}
\end{table}

\section{Pro-cyclicality through Volatility Binning} \label{app:vol_bins}

Another way to present the pro-cyclicality analysis of Section~\ref{ssec-volImpact}, more in line with the risk management point of view, may be through the volatility binning. Indeed, since we have seen that the SQP is under-estimated for low volatility and over-estimated for high one, we can show by how much the SQP is over- or underestimated for a hypothetical portfolio of a bank, depending on the volatility bins. 
To do so, we compute the average SQP ratios in each bin, choosing 5 and 10 uniform volatility bins.  Results are illustrated for 5 bins in Table~\ref{tbl-volBins}, and for 10 bins in Table~\ref{tbl-volBins10}.

In the case of 5 volatility bins we observe that the lowest bin underestimates on average the capital for next year by 35\%, whereas the highest one overestimates it by 22\%. 
From the lowest to the highest bin the underestimation transforms smoothly to an overestimation on average. For certain indices the behaviour in the three middle bins will be not monotone (CAN, FRA, USA) but for the lowest and highest bin we have the clear over- and underestimation as in the average behaviour.

The same holds true for 10 bins, but with increasing gaps between the extreme bins. On average we have in the lowest bin an overestimation of 44\% and in the highest bin an underestimation of 33\%. We observe that for all indices (with the exception of the USA)the under-/overrestimation, respectively, is highest for the lowest/ highest volatility bin.  

Those numbers are significant from the risk management prospective: It means increasing by more than a third (35 or 44) the capital amount in periods of low volatility, while decreasing it by more than a fifth (22 or 33) during crisis periods. 

\begin{table}[h]
\begin{center}
\parbox{440pt}{\caption{\label{tbl-volBins}\sf\small The average SQP ratios (defined in~\eqref{eq:VaRRatio}) within 5 uniform bins of volatility over the whole historical sample, for each index. In the last column, we present the average over all indices $\pm$ the standard deviation over the 11 displayed values.
}}
\vspace{-2ex}
\footnotesize
\begin{tabular*}{440pt}{c c c c c c c c c c c c}    \hline \\[-1.5ex]
AUS & CAN& FRA& DEU & ITA & JPN & NLD & SGP &SWE & GBR & USA & {\bf AVG ($\pm~\sigma$)}  \\[1ex] \hline\hline \\[.5ex]
1.21 & 1.23 & 1.35 & 1.35 & 1.36 & 1.64 & 1.36 & 1.30 & 1.38  & 1.33  & 1.31  & {\bf 1.35} $\pm$  0.11 \\ [1ex]
1.09 & 1.03 & 1.06 & 1.13 & 1.17 & 1.11 & 1.15 & 1.33 & 1.24  & 1.09  & 1.09  & {\bf 1.14}  $\pm$ 0.08 \\ [1ex]
1.07 & 1.20 & 1.12 & 1.10 & 1.22 & 0.95 & 1.11 & 0.98 & 1.11  & 1.04  & 1.01  & {\bf 1.08} $\pm$ 0.08 \\ [1ex]
0.96 & 0.99 & 0.99 & 1.11 & 0.94 & 0.93 & 1.04 & 0.82 & 0.86  & 1.03  & 1.11  & {\bf 0.98} $\pm$ 0.09 \\ [1ex]
0.77 & 0.87 & 0.78 & 0.74 & 0.68 & 0.79 & 0.73 & 0.83 & 0.79  & 0.81  & 0.74  & {\bf 0.78}  $\pm$ 0.05 \\ [2ex]
\hline\\
\end{tabular*}
\end{center}
\end{table}

\begin{table}[h]
\begin{center}
\parbox{440pt}{\caption{\label{tbl-volBins10}\sf\small The average SQP ratios (defined in~\eqref{eq:VaRRatio}) within 10 uniform bins of volatility over the whole historical sample, for each index. In the last column, we present the average over all indices $\pm$ the standard deviation over the 11 displayed values.
}}
%\vspace{-12ex}
\footnotesize
\begin{tabular*}{440pt}{c c c c c c c c c c c c}    \hline \\[-1.5ex]
AUS & CAN& FRA& DEU & ITA & JPN & NLD & SGP &SWE & GBR & USA & {\bf AVG ($\pm~\sigma$)}  \\[1ex] \hline\hline \\[.5ex]
1.29 &1.32 &1.42 &1.37 &1.40 &1.94 &1.37 &1.43& 1.54&  1.43&  1.28& {\bf 1.44} $\pm$  0.17 \\ [1ex]
1.12 &1.15& 1.29& 1.33& 1.32& 1.33& 1.35& 1.16 &1.22&  1.23&  1.34& {\bf 1.26} $\pm$  0.08 \\ [1ex]
1.14 &0.96 &1.08 &1.20 &1.13 &1.17 &1.15 &1.46 &1.26 & 1.14 & 1.08 & {\bf 1.16} $\pm$  0.12 \\ [1ex]
1.04 &1.11 &1.03 &1.06 &1.21 &1.05 &1.14 &1.20 &1.22&  1.04&  1.10&  {\bf 1.11} $\pm$  0.07 \\ [1ex]
1.08 &1.08 &1.19 &1.12 &1.26& 1.02& 1.21 &1.10 &1.06&  1.10&  1.07& {\bf 1.12} $\pm$  0.07\\ [1ex]
1.06 &1.32 &1.06 &1.08 &1.18 &0.89 &1.00 &0.86 &1.15 & 0.98&  0.96& {\bf 1.05} $\pm$  0.13 \\ [1ex]
1.00 &1.21 &1.00 &1.18 &1.00 &0.94 &1.07& 0.78& 0.86&  1.09&  1.05& {\bf 1.02} $\pm$ 0.12 \\ [1ex]
0.93 &0.78 &0.98 &1.04 &0.88 &0.92 &1.01 &0.85 &0.86&  0.97&  1.18& {\bf 0.95} $\pm$  0.10 \\ [1ex]
0.77 &1.04 &0.94 &0.88 &0.73 &0.84 &0.87 &1.00 &0.86 & 0.84&  0.89&  {\bf 0.88} $\pm$  0.09 \\ [1ex]
0.77 &0.70 &0.62 &0.60 &0.63 &0.73 &0.58 &0.65 &0.72 & 0.78&  0.59& {\bf 0.67} $\pm$  0.07 \\ [2ex]
\hline\\
\end{tabular*}
\end{center}
\end{table}

\end{appendices}

\end{document}